\documentclass{aa}  

\usepackage{graphicx}
\usepackage{natbib}
\usepackage{array}
\usepackage{rotating}
\usepackage{txfonts}
\usepackage{multirow}
\newcommand{\cygc}{Cyg\,OB2\,\#5}
\newcommand{\cygh}{Cyg\,OB2\,\#8A}
\newcommand{\cygd}{Cyg\,OB2\,\#12}
\newcommand{\xmm}{{\sc XMM}\emph{-Newton}}
\newcommand{\sw}{{\it Swift}}

\newcommand{\loglxlb}{$\log[L_{\rm X}/L_{\rm BOL}]$}

\begin{document}

   \title{Wind collisions in three massive stars of Cyg\,OB2\thanks{Based on observations collected at the Observatoire de Haute Provence (OHP) as well as with \sw\ and \xmm.}}

   \author{Constantin Cazorla
          \and
          Ya\"el Naz\'e\thanks{Research associate FNRS.}
          \and
          Gregor Rauw
          }

   \institute{D\'epartement AGO, Universit\'e de Li\`ege, All\'ee du 6 Ao\^ut 17, B\^at. B5C, B4000-Li\`ege, Belgium
              \email{cazorla@astro.ulg.ac.be}
             }

   \date{Received August 26, 2013; accepted ..., 2013}

% \abstract{}{}{}{}{} 
% 5 {} token are mandatory
 
  \abstract
   {}
   {We wish to study the origin of the X-ray emission of three massive stars in the Cyg\,OB2 association: \cygc, \cygh, and \cygd. }
   {To this aim, dedicated X-ray observations from \xmm\ and \sw\ are used, as well as archival {\it ROSAT} and {\it Suzaku} data.}
   {Our results on \cygh\ improve the phase coverage of the orbit and confirm previous studies: the signature of a wind-wind collision is conspicuous. In addition, signatures of a wind-wind collision are also detected in \cygc, but the X-ray emission appears to be associated with the collision between the inner binary and the tertiary component orbiting it with a 6.7\,yr period, without a putative collision inside the binary. The X-ray properties strongly constrain the orbital parameters, notably allowing us to discard some proposed orbital solutions. To improve the knowledge of the orbit, we revisit the light curves and radial velocity of the inner binary, looking for reflex motion induced by the third star. Finally, the X-ray emission of \cygd\ is also analyzed. It shows a marked decrease in recent years, compatible with either a wind-wind collision in a wide binary or the aftermath of a recent eruption. }
   {}


   \maketitle
%
%________________________________________________________________

\section{Introduction}
Massive stars lose large amounts of material during their lifetime, in the form of winds. In systems composed of several massive stars, these winds collide, giving rise to a wind-wind collision (WWC). These collisions might imprint their signature throughout the whole electromagnetic spectrum: nonthermal synchrotron emission \citep[in the radio range, see][as well as in the gamma-ray range, see \citealt{ley10}]{van06,deb07}, periodic IR-dust emission \citep[e.g.,][]{tut99}, line profile changes in the optical \citep[e.g.,][]{rau01}, hard thermal X-ray emission \citep[e.g.,][]{naz12}. To ascertain the origin of such emissions, very high angular resolution is necessary to resolve the WWC region from the stars \citep{dou05,pit,zhe}, or monitoring is required as the emission associated with WWCs usually varies with orbital phase. This paper focuses on the second approach.

One source of variation is the changing WWC orientation around the orbit. If the winds are of equal strengths, the WWC zone is planar and halfway between the two massive objects; if the winds are different, the WWC zone is a cone wrapped around the star with the weaker wind \citep{ste92}. As the system rotates, the line-of-sight to this WWC region changes, modifying the observed radial velocity of the shocked wind, hence the associated line profiles \citep{san01,hen03}. The rotation also induces changes in absorption, as the line-of-sight crosses the wind of each component in turn: if the wind densities differ sufficiently, a modulation of the absorption is then detected. The extreme case concerns very asymmetric binaries composed of an O-star and a Wolf-Rayet \citep{wil95,fau11}.\\
\begin{table*}
\caption[]{Journal of observations (date of mid-exposure, duration, identifier, and phases for the known periods, see text for details).}
\label{journal}
\centering
\begin{tabular}{ccccccc}
\hline\hline
\multirow{3}{*}{Facility}&\multirow{3}{*}{JD $-$ 2400000}& \multirow{3}{*}{$\Delta t$(d)}&\multirow{3}{*}{Observation ID/Rev}& \multicolumn{3}{c}{Phases}\\
&&&&\multicolumn{2}{c}{\cygc }& \cygh \\
&&&& 6.6\,d & 6.7\,yrs & 21.9\,d\\
\hline
\multirow{7}{*}{\xmm}&53308.579 &0.242&0200450201/0896&0.354&0.784&0.534\\
 &53318.558&0.266&0200450301/0901&0.867&0.788&0.989\\
 &53328.543&0.290&0200450401/0906&0.380&0.792&0.445\\
 &53338.505&0.266&0200450501/0911&0.890&0.796&0.900\\
 &54220.355&0.368&0505110301/1353&0.541&0.157&0.152\\
 &54224.170&0.382&0505110401/1355&0.120&0.159&0.326\\
 &55738.254&0.344&0677980601/2114&0.591&0.778&0.437\\
 \hline
\multirow{6}{*}{\sw}&55571.619&0.214&00031904001&0.337&/&0.831\\
 &55655.836&0.288&00031904002&0.100&/&0.675\\
 &55700.082&0.218&00031904003&0.806&/&0.695\\
 &55743.839&0.616&00031904004&0.438&/&0.692\\
 &55842.169&1.020&00031904005&0.340&/&0.181\\
 &56380.738&0.680&00032767001$+$2&0.965&0.040&0.764\\
 \hline
\multirow{6}{*}{\it ROSAT}&48368.064&0.274&200109&0.570&0.764&/\\
 &49109.312&4.554&900314&/&0.067&/\\
 &49107.150&0.231&900314 : 1/4&0.586&/&/\\  
 &49109.465&0.621&900314 : 2/4&0.937&/&/\\
 &49110.275&1.094&900314 : 3/4&0.060&/&/\\
 &49111.201&0.952&900314 : 4/4&0.200&/&/\\ 
 \hline
{\it Suzaku}&54453.965&0.925&402030010&/&0.252&/\\
 \hline
\end{tabular}
\end{table*}
\indent A second source of variations is linked to the changing distance between the components in eccentric systems. The behavior depends on the value of the cooling parameter, i.e., the ratio between the cooling time of the shocked gas and the escape time. It can be expressed as $\chi = \frac{v_{8}^{4}\,d}{\dot{M}_{-7}}$, where $v_{8}$ is the wind velocity in units of 1000 km s$^{-1}$, $d$ the distance to the contact discontinuity expressed in units of 10$^{7}$ km, and $\dot{M}_{-7}$ the mass loss rate in units of 10$^{-7}$ M$_{\odot}$ an$^{-1}$ \citep{ste92}. If $\chi\ll 1$, the gas cools very quickly and the collision is considered to be {\it radiative}. In this case, the X-ray emission should follow $L_{\rm X} \propto \dot{M}v^{2}$, where $v$ is the pre-shock wind velocity, and the shocked plasma temperature might be lower at periastron if the winds are still accelerating when they collide. This situation regularly occurs in short-period O+O 
systems, but might also happen in long-period systems comprising slow or dense winds, as $\chi$ is, in fact, inversely related to the wind number density at the shock. An example of such radiative collisions can be found in HD\,152248 \citep{san04}. If $\chi\ge 1$, the gas does not easily cool and the collision is then adiabatic. In this case, the properties of the WWC zone mainly depend on the wind density: the X-ray emission is expected to vary as the inverse of the separation between the components. This case is relevant for most systems with orbital periods longer than a few days \citep{ste92} as in Cyg\,OB2\,\#9\ \citep{naz12} and WR25 \citep{gos07}. Note that radiative cooling might become significant, even in adiabatic collisions, in regions away from the line of centers, where the angle of collision is oblique, or at certain phases (such as periastron).\\  
\indent Massive stars are intrinsic sources of X-rays, with a tight relation between the bolometric and high-energy luminosities \citep[\loglxlb$\sim-7$, e.g.,][]{ber97}. Two decades ago, some massive binaries were detected to be brighter in the X-ray domain, the additional emission being attributed to WWCs \citep{pol87,chl91}. In recent years, however, the paradigm shifted, as many massive, O+OB binaries were not found to be overluminous \citep{pit00,osk05,naz09,naz11}. The difference in luminosity between binaries and single objects is generally very small, at best \citep{naz13}. Only few massive binaries appear strongly overluminous in the X-ray range, and even fewer have been monitored in detail (for a review, see \citealt{gue09}), e.g., HD\,93403 \citep{rau02}, or WR22 \citep{gos09}. These few cases are, however, the only ones to provide a testbed for theoretical WWC models and constrain the (still debated) stellar wind properties: finding new ones or clarifying the properties of known WWCs 
is, therefore, highly important.\\
\indent The first massive stars detected in the X-ray range belonged to the Cyg\,OB2 association \citep{har79}. The four brightest sources were associated with \cygc, \cygh, Cyg\,OB2\,\#9, and \cygd. Over the years, these objects have been sporadically observed at high energies, but a better sampling of the orbital cycle is needed. The first X-ray campaigns revealed changes in \cygh\ \citep{deb07,blo10} and Cyg\,OB2\,\#9 \citep{naz12}. This paper provides further results from X-ray monitoring of \cygh, \cygc, and \cygd. Section \ref{section_2_observations} presents the observations, Sections \ref{section_cygob28A}, \ref{section_cygob25}, and \ref{section_cygob212} analyze the situation of each of these three massive stars in turn, and Section \ref{section_conclusions} summarizes our findings and concludes this paper.

\section{Observations}
\label{section_2_observations}
For our study, we rely on the same \xmm\ and \sw\ datasets as used for Cyg\,OB2\,\#9 in \citet{naz12}, with the addition of one \sw\ XRT exposure taken in March 2013. To ensure homogeneity and the use of the latest calibration, these X-ray data were processed again, using SAS v12.0.0. for \xmm\ data and HEASOFT v6.13 for \sw\ data, following the recommendations of the respective instrument teams (European Space Agency Science Operations Centre (ESA SOC) and UK \sw\ center). We also used archival {\it Suzaku} and {\it ROSAT} data of Cyg\,OB2. Table \ref{journal} provides the identifier of these observations, as well as their date. 

For \xmm\ observations, a source detection algorithm ({\it edetect\_chain}) was used to derive the position of our targets in each exposure. To ensure homogeneity, the shapes and relative positions of source and background regions remain the same for all seven \xmm\ datasets. This is a complex task as the position angle and center of the field-of-view change for each exposure\footnote{For \cygh, a unique background region for all pn observations could not be defined. We used two regions, one for the first four \xmm\ observations, and one for the three remaining observations.}. When possible, we then extract source events in circular regions of radii 50'', 23'', and 15''\footnote{We note however that a circular region radius of 8'' for the first pn observation (ObsID 0200450201) had to be considered.} (for MOS1, MOS2, and pn detectors, respectively) for \cygc, 14'', 46'', and 16'' for the same three detectors for \cygd, and 26'' for \cygh\ (same region for all EPIC detectors). We selected circular 
(for MOS) and polygonal (for pn) background regions, as close as possible to the targets and devoid of other sources. We calculated individual response matrices for each target and each observation (tasks {\it rmfgen, arfgen}). We performed pile-up checks, showing no impact for our targets. The last four \xmm\ observations were affected by episodes of soft proton flares, which we discarded. Note, however, that the results of spectra fitting remain the same whether we keep or cut the time intervals affected by flares. For \sw\ observations, we extracted source events in circular regions centered on the Simbad coordinates of the targets and with radii 21'' for \cygc\ and \cygd, and 11'' for \cygh. A polygonal background region, common to all targets, was used: it is located within the trapezium formed by the targets and Cyg\,OB2\,\#9 and it is as large as possible, to ensure good statistics on the background and to avoid any localized background variation. For the \sw\ observations, we used 
the response 
matrix file 
provided by the \sw\ team (swxpc0to12s6\_20010101v013.rmf), but calculated an individual ancillary response file specifically for the targets using the task {\it xrtmkarf}, with the inclusion of an exposure map so that we take bad columns into account. For {\it ROSAT} observations, we extracted source events in circular regions centered on the Simbad coordinates of the targets and with radii 61'', 27'', and 42'' for \cygc, \cygh, and \cygd, respectively; we used as backgrounds a nearby circular region of radius 39" for \cygh\ and annular regions of outer radii of 97'' and 84'' (the inner radii being equal to the radii of the source regions) for \cygc\ and \cygd, respectively. The archival Redistribution Matrix Files (RMFs) for the Position Sensitive Proportional Counter B (PSPCB) (pspcb\_gain1\_256.rmf and pspcb\_gain2\_256.rmf for the first and second observations, 
respectively) were used, and we calculated individual ARFs for each target and each exposure (task {\it pcarf}). For the {\it Suzaku} observation, source events were extracted in circular regions centered on the 
Simbad coordinates of the targets with radii 100'' and 90'' for \cygc\ and \cygd, respectively, while we used nearby circular regions with 100'' radii for background. Response matrices were calculated for the source using online calibration files, as recommended for {\it Suzaku} spectra. Data from both $3\times 3$ and $5\times 5$ modes and from all available X-ray Imaging Spectrometer (XIS) chips (0, 1, and 3) were then combined.\\
\indent Some observations were grouped or split, depending on the period of the system under scrutiny. For \cygc, the first five \sw\ observations, taken a few months apart, were grouped when considering the long 6.7\,yr period, as they present similar phases in this case (see Section \ref{section_cygob25}), and the second {\it ROSAT} observation (ObsID 900314) was split into four exposures when considering the short 6.6\,d period as that observation covers a large part of the 6.6\,d orbit (see Table \ref{journal}). For \cygh, we grouped the second, third, and fourth \sw\ observations as they correspond to similar phases in the 21.9\,d period of the system. For \cygd, we grouped the first five \sw\ observations since the main variations occur on long timescales.\\
\indent Finally, we fitted all extracted spectra within Xspec v12.8.0 (using $apec$ v2.0.1). A combination of models for optically thin thermal plasma absorbed by interstellar and wind material\footnote{Although this contribution comes from ionized material, a neutral absorption component was used to represent it. The difference in absorptions by neutral and ionized material occurs below 1\,keV, where there is no usable data due to the high interstellar absorption in Cyg\,OB2.} was used: two to three thermal emission components were necessary to obtain a good fit (see below). Fits to the \xmm\ data were first performed allowing all the parameters to vary, and when a parameter was seen to remain constant, it was fixed; for the \sw, {\it ROSAT} and {\it Suzaku} spectra, it was often necessary to fix as many parameters as possible, to avoid erratic variations (see below for details).

\section{The object \cygh }
\label{section_cygob28A}
The object \cygh\ is an O6If + O5.5III(f) binary with a period of 21.9\,d \citep{deb04}. It is also a nonthermal radio emitter, indicating that a WWC occurs in the system. Phase-locked modulation of the radio emission was reported by \citet{blo10}. The same authors also derived improved orbital and physical parameters for the stars. The bright X-ray emission also displays phase-locked variability \citep{deb06}, but the radio and X-ray emissions show anticorrelated behaviors \citep{blo10}, as the formation regions are different (closer to the apex of the shock cone for X-rays).

\begin{figure}
   \centering
  
  \includegraphics[scale=0.3]{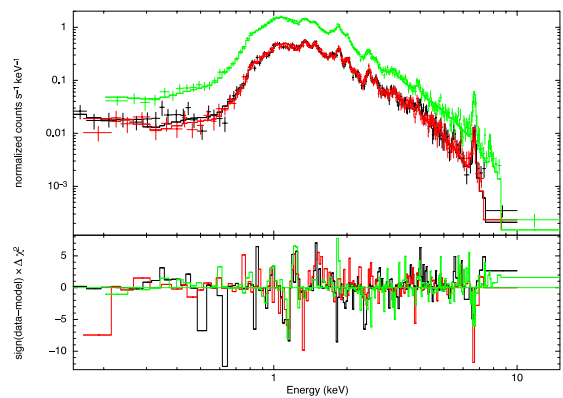}
    \caption{\cygh\ spectra acquired with the MOS1 (black), MOS2 (red), and pn (green) detectors in November 2004 (Rev. 0906) along with the best-fit model and its residuals.}
         \label{cyg8Aspec}
   \end{figure}

We used \sw\ data along with an additional \xmm\ observation to increase phase coverage. We used an interstellar absorbing column of $0.91\times10^{22}$\,cm$^{-2}$ (corresponding to $E(B-V)=1.56$ of \citealt{weg03} when using the \citealt{boh78} conversion ratio) and the ephemeris of \citet{deb04}. The associated phases of the observations are listed in Table \ref{journal}. First, we performed spectral fitting of the sole \xmm\ data (Fig. \ref{cyg8Aspec}). Three thermal components were necessary to achieve a good fit, and it should be noted that the temperatures at apastron ($\phi=0.5$) are slightly higher than at periastron, which is consistent with the results of \citet{blo10}. This can be explained by the larger wind speeds at apastron, as the winds have more time to accelerate before they collide. However, this increase is within the error bars, hence is not formally significant \citep[as in][]{blo10} so that the derived temperatures can be considered constant within the uncertainties. We, therefore, 
fixed these temperatures and fitted all \xmm\ spectra again. The normalization factor of the first thermal component appears slightly lower at phase $\phi=0.5$ (i.e., apastron), but this is again not significant considering the uncertainties. This implies that most of this component is born in the winds of the stars, not at the WWC zone. We thus fitted all spectra again with temperatures and first normalization fixed to the average value from \xmm\ fits (Table \ref{fit8}). We note that this does not change the results significantly compared to fully-free fitting.

\begin{figure}
   \centering
   \includegraphics[width=8cm,bb=20 20 520 400, clip]{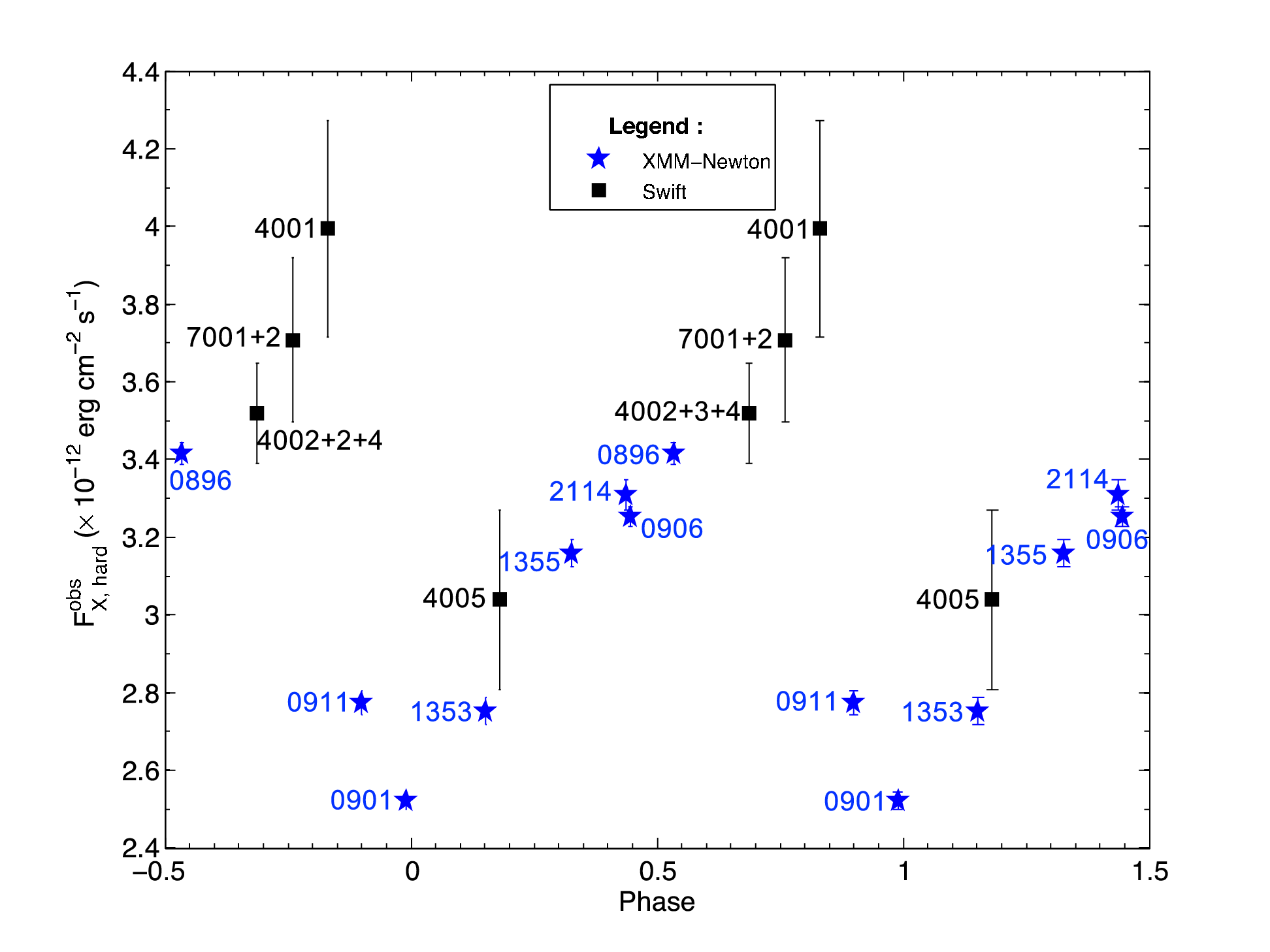}
   \includegraphics[width=8cm,bb=20 20 520 400, clip]{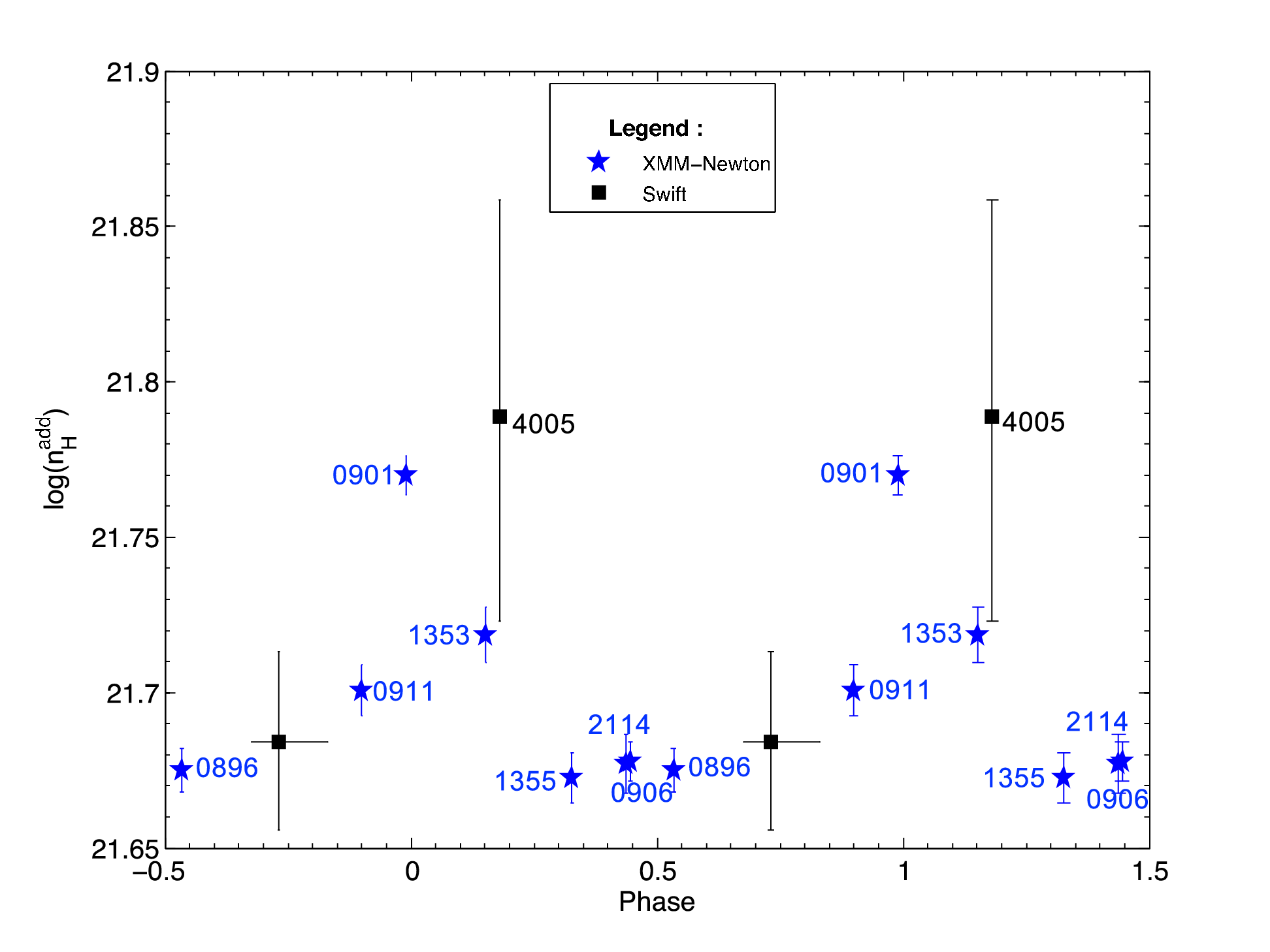}
   \includegraphics[width=8cm,bb=20 20 520 400, clip]{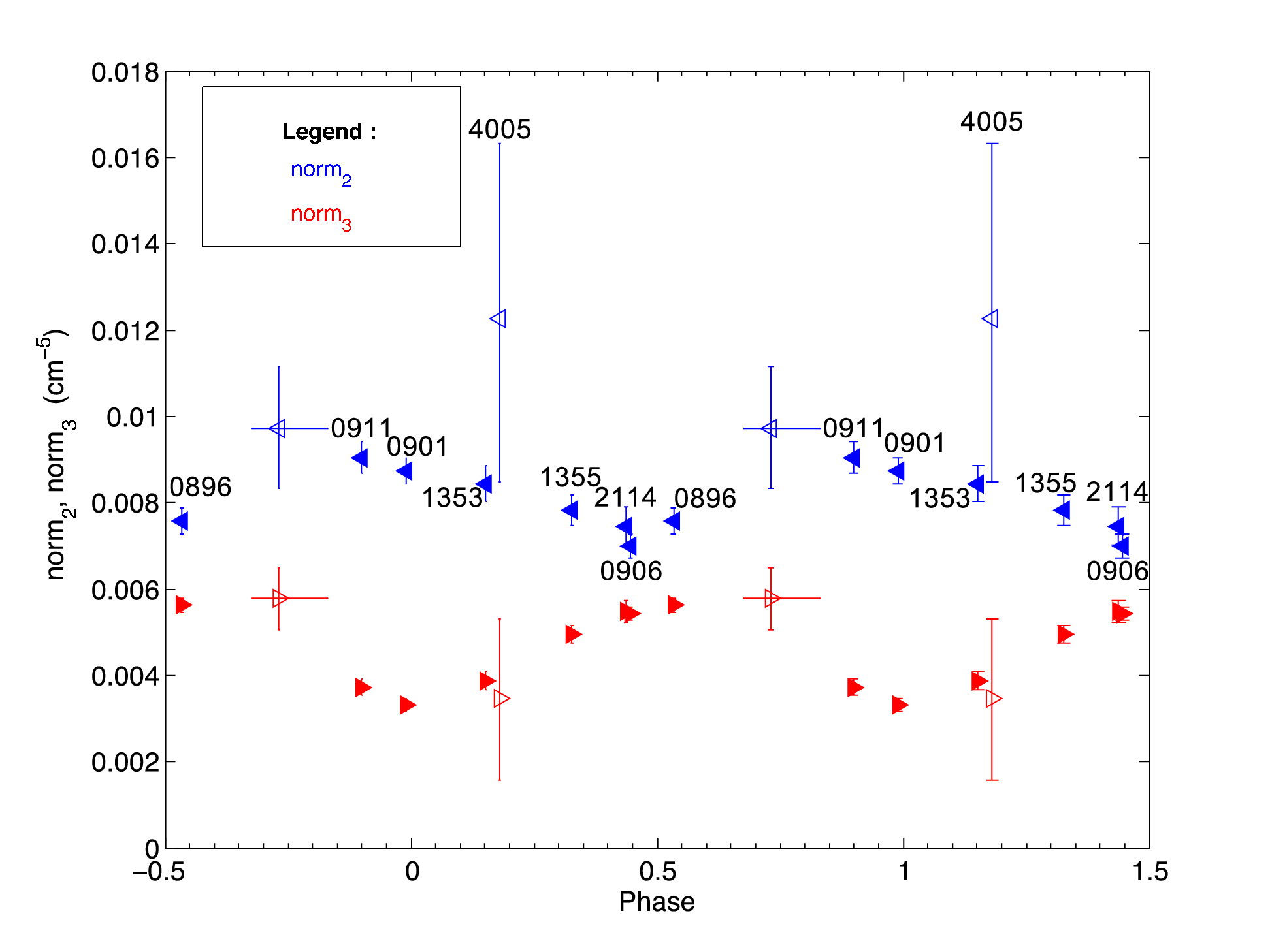}
      \caption{Evolution with phase of the hard X-ray flux (top), the absorption (middle) and the normalization factors (bottom) for \cygh. In the bottom panel, filled and empty symbols represent \xmm\ and \sw\ data, respectively. Note that for the middle and bottom panels, all \sw\ data but those of 00031904005 were grouped to increase signal-to-noise, the length of the bin representing the phase interval covered by the data.}
         \label{cyg8}
   \end{figure}

Looking at the results, several conclusions can be drawn. The additional absorption is maximum at periastron and minimum at apastron (Fig. \ref{cyg8}), in agreement with the findings of \citet{deb06}. This is linked to the system's orientation: the primary star, whose wind is strongest, is in front of the secondary star near periastron, resulting in a larger additional absorption; on the contrary, the secondary star is in front of the primary star near apastron, resulting in a smaller absorption.

  
      \begin{figure}
   \centering
   \includegraphics[width=9cm,bb=40 180 570 440, clip]{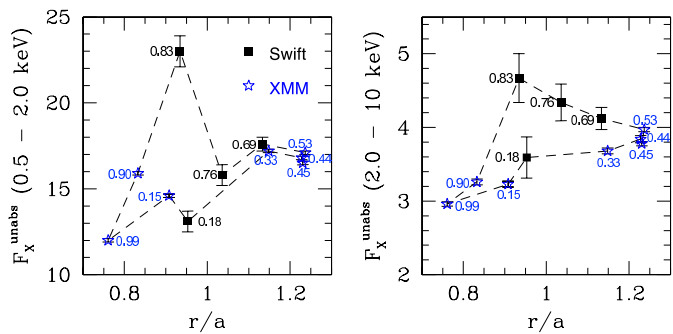}
      \caption{Evolution with relative separation ($r/a$ with $a$ the semi-major axis) of the soft and hard X-ray fluxes for \cygh. }
         \label{hyster}
   \end{figure}

\begin{sidewaystable}
\caption{Results of the X-ray spectral fitting for \cygh.}
\label{fit8}
\centering
\begin{tabular}{ccccccccccccccccc }
\hline\hline
\multirow{3}{*}{Facility} & \multirow{3}{*}{Observation ID} &  \multirow{2}{*}{$n_{\rm H}^{\text{add}}$} &  \multirow{2}{*}{\textit{norm}$_{2}$} &  \multirow{2}{*}{\textit{norm}$_{3}$} & \multicolumn{3}{c}{$F_{X}^{\text{obs}}$} &  \multicolumn{3}{c}{$F_{X}^{\text{unabs}}$} & \multirow{3}{*}{$\chi^{2}_{\nu}$ (d.o.f.)} \\
 &&&&&\multicolumn{3}{c}{(10$^{-12}$ erg cm$^{-2}$ s$^{-1}$)}&\multicolumn{3}{c}{(10$^{-12}$ erg cm$^{-2}$ s$^{-1}$)}&\\
 &&($10^{22}$ cm$^{-2}$)&($10^{-3}$ cm$^{-5}$)&($10^{-3}$ cm$^{-5}$)&Total&Soft&Hard&Total&Soft&Hard&\\
\hline
\multirow{7}{*}{\xmm}&0200450201&{0.47$_{-0.01}^{+0.01}$}& {7.57$_{-0.31}^{+0.31}$}& {5.63$_{-0.17}^{+0.17}$}&{6.24$\pm0.03$}& {2.83$\pm0.01$}& {3.42$\pm0.03$}& {21.1$\pm0.1$}& {17.1$\pm0.1$}& 3.97$\pm0.03$&{1.37\hspace{1mm}(551)}\\ 
 &0200450301&{0.59$_{-0.01}^{+0.01}$}& {8.74$_{-0.30}^{+0.30}$}& {3.32$_{-0.15}^{+0.15}$}& {4.83$\pm0.02$}& {2.31$\pm0.01$}& {2.52$\pm0.02$}& {15.0$\pm0.1$}& {12.0$\pm0.1$}& {2.96$\pm0.03$}& {1.30\hspace{1mm}(504)}\\
 &0200450401&{0.48$_{-0.01}^{+0.01}$}& {7.00$_{-0.27}^{+0.28}$}& {5.43$_{-0.15}^{+0.15}$}& {5.94$\pm0.02$}& {2.69$\pm0.01$}& {3.25$\pm0.03$}& {20.2$\pm0.1$}& {16.5$\pm0.1$}& {3.78$\pm0.03$}& {1.39\hspace{1mm}(568)}\\
 &0200450501&{0.50$_{-0.01}^{+0.01}$}& {9.04$_{-0.37}^{+0.37}$}& {3.72$_{-0.19}^{+0.19}$}& {5.48$\pm0.04$}& {2.71$\pm0.02$}& {2.77$\pm0.03$}& {19.2$\pm0.1$}& {15.9$\pm0.1$}& {3.26$\pm0.04$}& {1.37\hspace{1mm}(477)}\\
 &0505110301&{0.52$_{-0.01}^{+0.01}$}& {8.44$_{-0.41}^{+0.41}$}& {3.88$_{-0.22}^{+0.21}$}& {5.30$\pm0.03$}& {2.55$\pm0.02$}& {2.75$\pm0.04$}& {17.8$\pm0.1$}& {14.6$\pm0.1$}& {3.23$\pm0.04$}& {1.09\hspace{1mm}(452)}\\
 &0505110401& {0.47$_{-0.01}^{+0.01}$}& {7.82$_{-0.35}^{+0.35}$}& {4.95$_{-0.19}^{+0.19}$}&{5.95$\pm0.03$}& {2.79$\pm0.02$}& {3.16$\pm0.03$}& {20.9$\pm0.1$}& {17.2$\pm0.1$}& {3.68$\pm0.04$}& {1.38\hspace{1mm}(517)}\\
 &0677980601& {0.48$_{-0.01}^{+0.01}$}& {7.45$_{-0.44}^{+0.44}$}& {5.48$_{-0.25}^{+0.25}$}&{6.11$\pm0.04$}& {2.80$\pm0.02$}& {3.31$\pm0.04$}& {20.7$\pm0.1$}& {16.8$\pm0.1$}& {3.84$\pm0.05$}& {1.39\hspace{1mm}(188)}\\
\hline
\multirow{5}{*}{\sw}&00031904001& {0.41$_{-0.07}^{+0.08}$}& {9.08$_{-3.46}^{+3.82}$}& {6.37$_{-2.00}^{+1.92}$}& {7.43$\pm0.26$}& {3.43$\pm0.14$}& {3.99$\pm0.28$}& {27.2$\pm1.0$}& {22.5$\pm0.9$}& {4.67$\pm0.33$}& {0.92\hspace{1mm}(68)}\\
&00031904002--4&{0.49$_{-0.04}^{+0.04}$}& {9.47$_{-1.76}^{+1.84}$}& {5.26$_{-0.92}^{+0.90}$}&{6.53$\pm0.13$}& {3.01$\pm0.07$}& {3.52$\pm0.13$}& {21.8$\pm0.4$}& {17.6$\pm0.4$}& {4.12$\pm0.15$}& {1.24\hspace{1mm}(172)}\\
&00031904005&{0.61$_{-0.09}^{+0.10}$}& {12.3$_{-3.8}^{+4.1}$}& {3.47$_{-1.90}^{+1.85}$}&{5.76$\pm0.23$}& {2.72$\pm0.13$}& {3.04$\pm0.23$}& {16.7$\pm0.7$}& {13.1$\pm0.6$}& {3.59$\pm0.28$}& {0.97\hspace{1mm}(53)}\\
&00032767001$+$2&{0.54$_{-0.06}^{+0.07}$}& {10.5$_{-2.9}^{+3.1}$}& {5.48$_{-1.60}^{+1.56}$}&{6.69$\pm0.20$}& {2.99$\pm0.11$}& {3.71$\pm0.21$}& {20.2$\pm0.6$}& {15.8$\pm0.6$}& {4.34$\pm0.25$}& {1.13\hspace{1mm}(83)}\\
& All but 4005 & 0.48$_{-0.03}^{+0.03}$ &  9.72$_{-1.38}^{+1.42}$ & 5.78$_{-0.72}^{+0.71}$ & 6.91$\pm0.10$ & 3.13$\pm0.06$ & 3.78$\pm0.11$ & 22.6$\pm0.3$ & 18.2$\pm0.3$ & 4.41$\pm0.12$ & 1.25 (225)\\
\hline
\end{tabular}
\tablefoot{The fitted model has the form $wabs*phabs*(apec+apec+apec)$, with the first absorption fixed to $0.91\times10^{22}$\,cm$^{-2}$, the temperatures fixed to 0.23, 0.93, and 2.0\,keV, and the normalization factor of the first thermal component fixed to 0.0706\,cm$^{-5}$. The lower and upper limits of the 90\% confidence intervals on spectral parameters can be derived from the indices and exponents, respectively, while the relative $\pm1\sigma$ errors on fluxes correspond to the relative $\pm1\sigma$ errors on count rates. Normalization factors are defined as $[10^{-14} / (4\,\pi\,D^{2})]\,\int n_{\rm e}\,n_{\rm H}\,dV$, where $D$ is the source distance, $n_{\rm e}$ and $n_{\rm H}$ the electron and hydrogen number densities, respectively. The total, soft, and hard energy bands correspond to the 0.5--10.0\,keV, 0.5--2.0\,keV, and 2.0--10.0\,keV intervals, respectively.}
\end{sidewaystable}

The normalization factors of the second and third thermal components vary in antiphase (Fig. \ref{cyg8}): the second normalization appears minimum at apastron, while the third one is maximum at that phase. This difference in behavior might appear puzzling, but is, in fact, normal in view of the formation zones of the thermal components. The hardest X-rays come from the head-on collision between the winds, which occurs at the apex of the shock cone, near the line joining the centers of the two stars. The variation of the normalization factor associated with the hard component is thus linked to the peculiarities of the WWC: in radiative systems such as \cygh, the collision occurs at higher speeds at apastron, when the winds have enough space to accelerate to higher speeds. The shock is thus stronger at apastron, explaining the increase in hard emission. The softest X-rays come from regions further down the shock cone, which are less sensitive to velocity differences, as the shock already occurs there 
at reduced speeds. The changes in the normalization factors associated with these lower temperatures (though slight for the first normalization) are rather dominated by variations in plasma density: the density, hence the emission, is higher when stars are close to each other.\\ 
\indent Combining the three components, the global flux indeed also varies (Fig. \ref{cyg8}). The particularity is that its peak does not occur at apastron or periastron, but at an intermediate phase of $\phi\sim 0.8$ \citep[this confirms the preliminary results of][]{deb06}. This probably stems from the behavior of the normalization factors: at that phase, the second normalization is already high, while the third normalization is not yet minimum (Fig. \ref{cyg8}). Hydrodynamic simulations of the WWC in \cygh\ aimed at reproducing the X-ray spectra at a few phases could reproduce the lower flux at periastron ($\phi$=0), but could not find a clear peak at that intermediate phase. Rather, a flux close to, but still lower than, the apastron flux ($\phi$=0.5) was found \citep{deb06}. However, more sophisticated models by \citet{pit10} yield somewhat different results. They reveal that the X-ray emission could be asymmetric around periastron (or apastron) in eccentric systems, 
as the emission properties at a 
given 
phase depend 
on the properties of the plasma created at earlier phases. Though the systems modeled by \citet{pit10} are different from \cygh\ (the only eccentric model has a shorter period and involves two identical main sequence objects, therefore, the shocked winds have different radiative timescales than here), a comparison with our observation reveals interesting similarities. In particular, an asymmetry is observed in the variation of the normalization factors and in the fluxes of \cygh. Once the fluxes are plotted as a function of separation (Fig. \ref{hyster}), as in Fig. 19 of \citet{pit10}, the predicted hysteresis behavior is clearly seen: the emission is harder as the stars get closer than when the stars separate. However, the models predict the largest hysteresis near periastron, where the properties of the shocked plasma rapidly change, but this is not observed: the largest variation occurs in between apastron and periastron. Definitely, a full hydro model, sampling the whole orbit, is now needed.

\section{The object \cygc }
\label{section_cygob25}
The object \cygc\ has long been known to be an eclipsing binary, but it was recently found to be a quadruple system \citep{ken10}. The core of \cygc\ is a short-period (6.6\,d), eclipsing binary composed of an O6.5--7I and an OB--Ofpe/WN9 transition object. \citet{lin09} showed that the stars are in a contact configuration, and that the secondary hemisphere facing the primary is hotter and brighter. Phases derived from \citet{lin09} are given in Table \ref{journal} for each observation. The system is a nonthermal radio emitter, which is an indication of the presence of a WWC. Strangely, the radio emission of the binary is not modulated with the 6.6\,d period, but with a period of about 6.7\,yrs. \citet{ken10}, therefore, deduced that a third star exists in the system, orbiting the binary in a 6.7\,yr orbit. This third component can be associated with a late O/early B-type star \citep{ken10}. These authors proposed several possible orbital solutions for the tertiary orbit that could explain the radio 
emission. The ephemeris of their favored model ($s=0$) is used for the phases shown in Table \ref{journal}, associated with the long period. Further away, at 0.9'' to the NE, a visual companion with early B-type is located: it is the fourth component of the complex \cygc\ system \citep{con97}. The orbital period of this fourth component around the triple system has been estimated at 9200\,yrs for a distance of 1.7 kpc \citep{lin09}. \citet{ken10} have detected a second WWC zone near this star, a nonthermal emission with the typical ``crescent moon'' shape associated with WWC shock cones. 

The X-ray emission of \cygc\ is somewhat brighter than usual for massive stars (\loglxlb$\sim -6.4$), but it was not known to vary much \citep{lin09}. In particular, no clear 6.6\,d modulation was seen in the first six \xmm\ observations \citep{lin09}, so that the origin of the X-ray emission remains unclear, especially in view of the new results on the system's composition.

\begin{figure}
  \centering
  
    \includegraphics[scale=0.3]{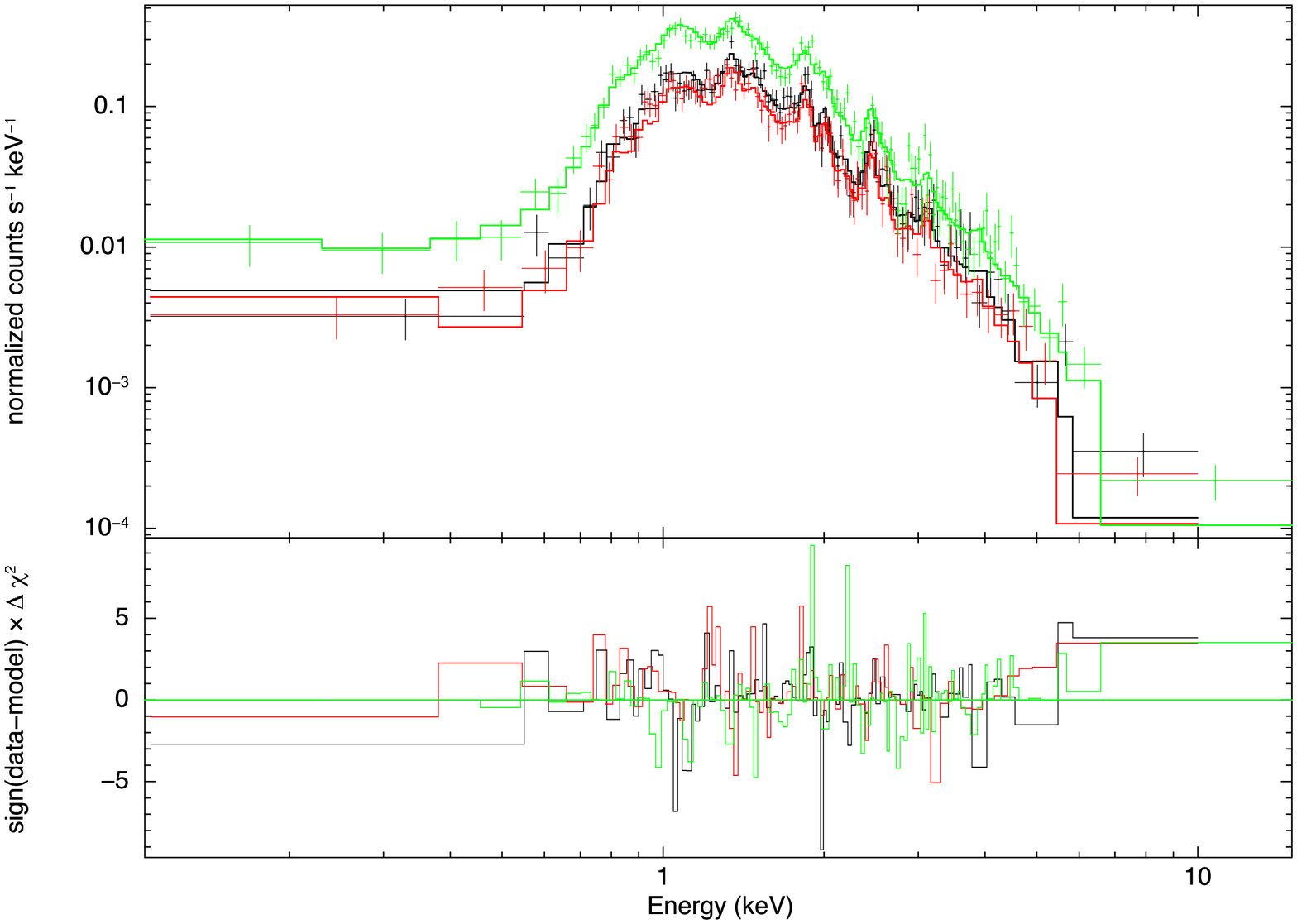}
      \  \caption{\cygc\ spectra acquired with the MOS1 (black), MOS2 (red) and pn (green) detectors in November 2004 (Rev. 0911) along with the best-fit model and its residuals.}
         \label{cyg5spec}
   \end{figure}

Our dataset includes much more data than were available to \citet{lin09}, enabling us to revisit the properties of \cygc. We used an interstellar absorbing column of $1.14 \times 10^{22}$\,cm$^{-2}$ (corresponding to $E(B-V)=1.96$), the quadratic ephemerides for the binary of \citet{lin09}, and the preliminary ephemeris for the third star of the favorite model of \citet{ken10}. The associated phases of the observations are given in Table \ref{journal}. As for \cygh, we first performed spectral fitting of the \xmm\ data only (Fig. \ref{cyg5spec}). As frequently happens when fitting X-ray spectra of massive stars, two sets of temperatures provide equally good fits: 0.2+1.2\,keV and 0.7+1.9\,keV. As the results of the fluxes and variations of the parameters are similar\footnote{The main differences between the two models are (1) the additional absorptions ($\sim0.6$ and $\sim0.3 \times 10^{22}$\,cm$^{-2}$ for the lower temperature pair and the larger temperature pair, respectively) and (2) 
the ratios 
between the two normalization factors ($\sim$14.9 and $\sim$4.0 for the lower temperature pair and the larger temperature pair, respectively).}, in the following we will only discuss the first solution. The derived temperatures can be considered constant within the uncertainties, so that we fixed them and fitted all \xmm\ spectra again (Table \ref{fit5}). The ratio between the two normalization factors remains similar, within the uncertainties, amongst the observations, so that we further fixed it to 14.9 for fitting the \sw, {\it Suzaku} and {\it ROSAT} spectra (Table \ref{fit5}).

\begin{figure}
   \centering
   \includegraphics[width=7.9cm,bb=14 8 530 403,clip]{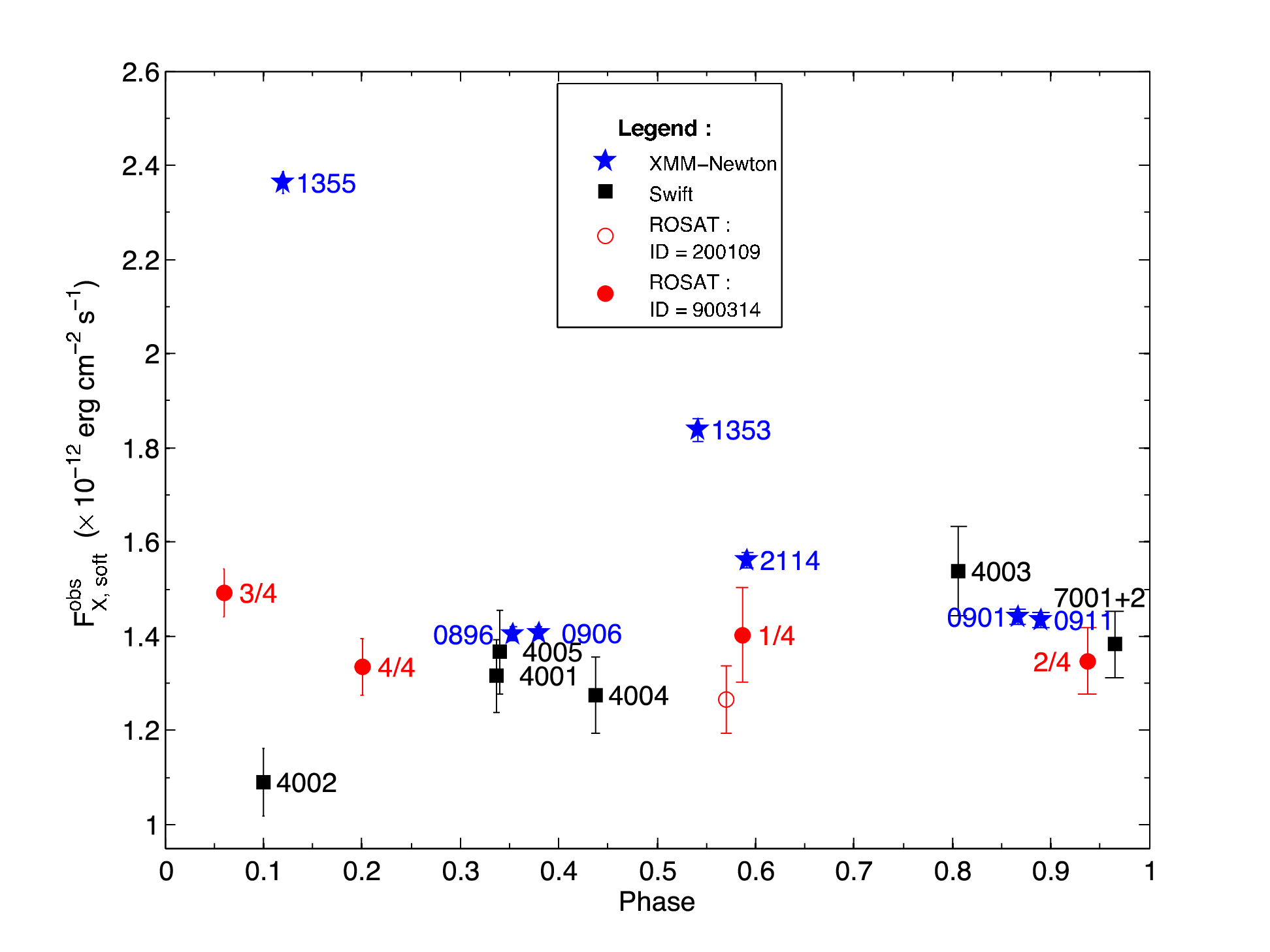}
   \includegraphics[width=7.9cm,bb=14 8 530 403,clip]{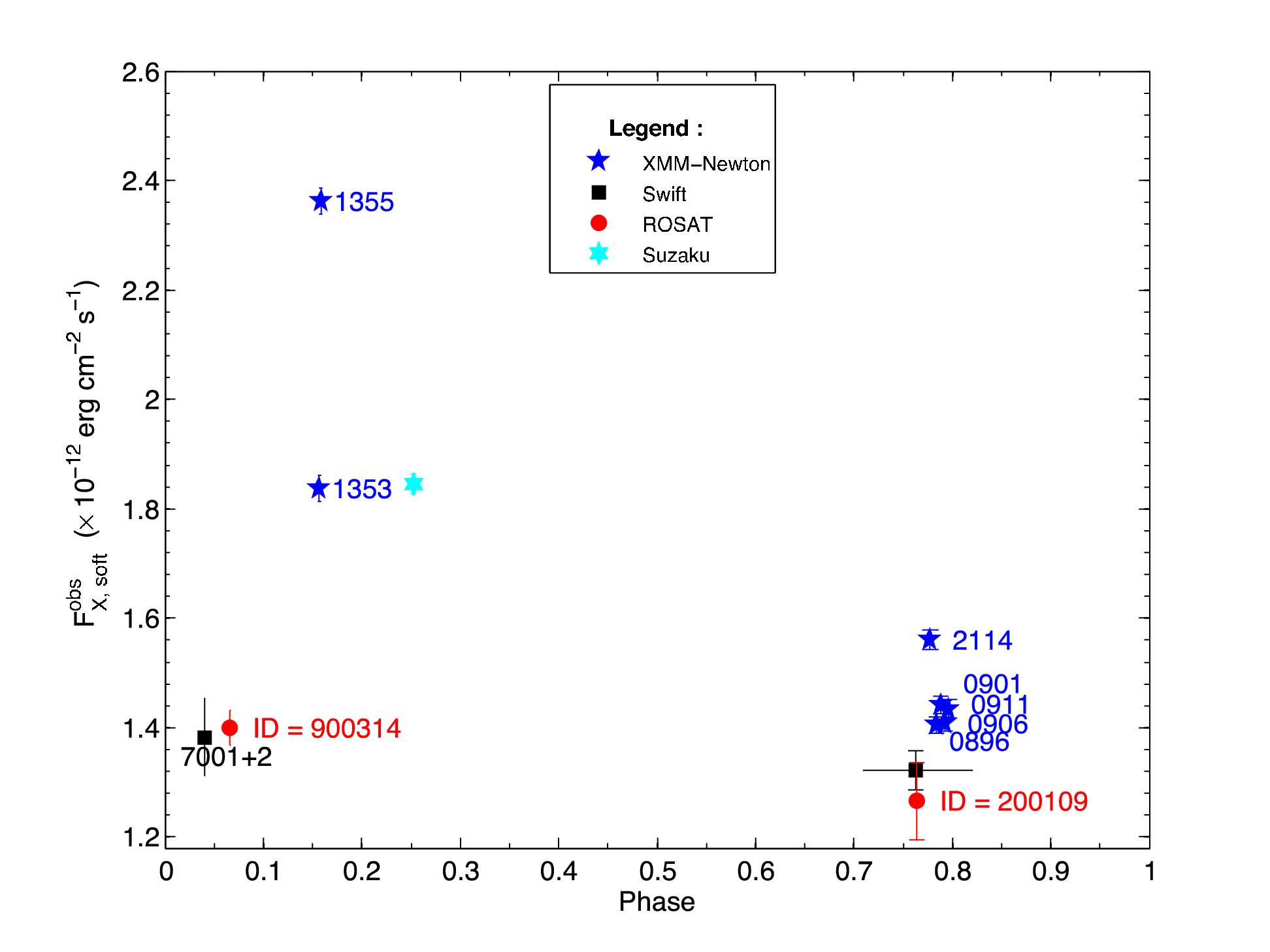}
   \includegraphics[width=7.9cm,bb=14 8 530 403,clip]{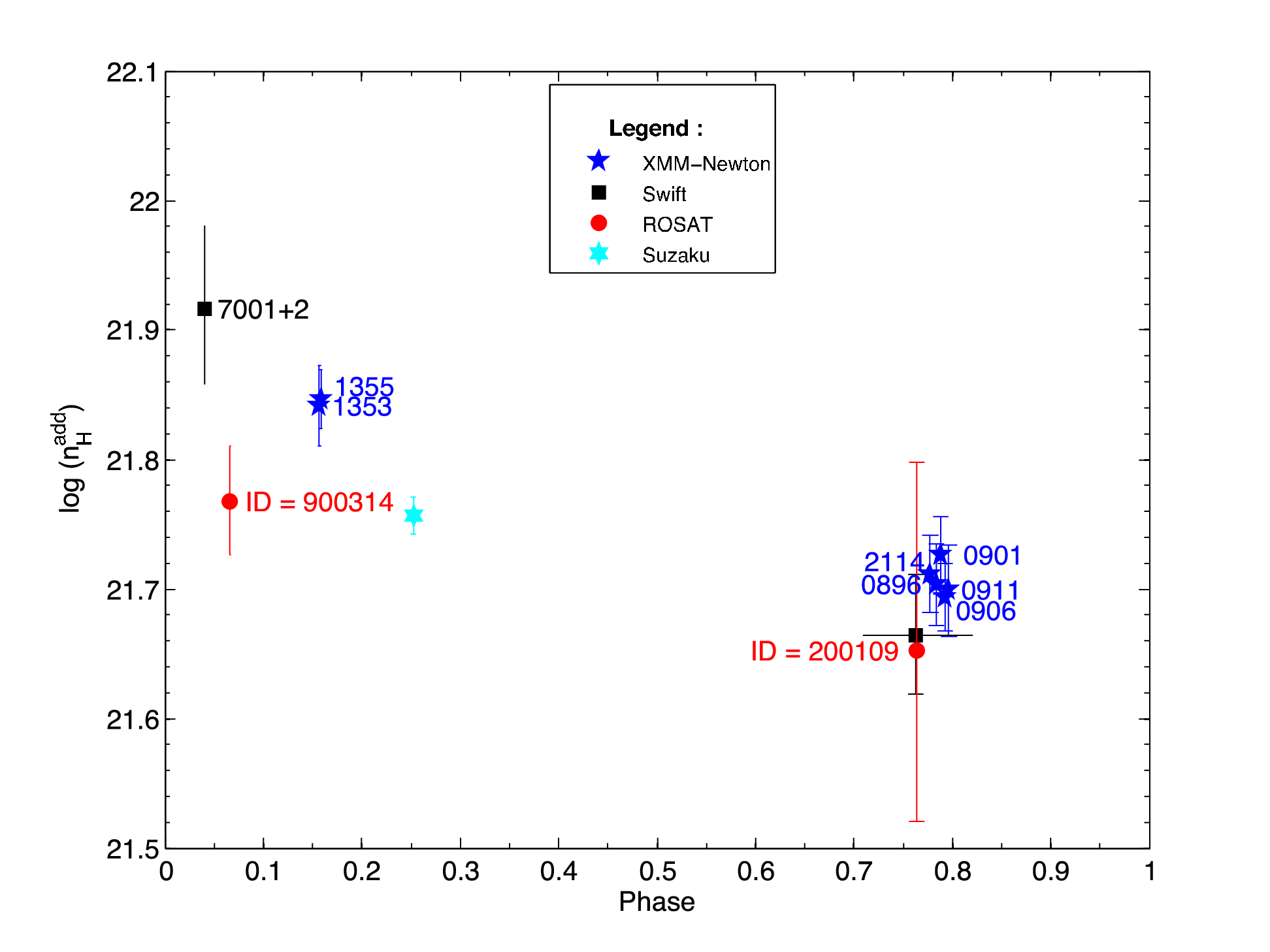}
      \caption{Evolution of the soft X-ray flux with phase considering two periods of \cygc: the short period of the binary (6.6\,d, top) and the long period of the triple system (6.7\,yrs, middle). The bottom panel shows the evolution with phase of the absorption, for the latter period only.}
         \label{flux5}
   \end{figure}

The new dataset covers the 6.6\,d orbit several times, but yields surprising results. Indeed, when plotted against the phase in the 6.6\,d period, the flux appears scattered (Fig. \ref{flux5}): the \xmm\ data of Rev. 1355 appears much brighter than the \sw\ data taken at a similar phase, and the \xmm\ data of Rev. 1353 appear much brighter than those of Rev. 2114 and some {\it ROSAT} observations (ObsIDs 200109 and first part of 900314), despite a similar phase. The modulation of the X-ray properties is thus not associated with the binary period. We, therefore, tested the data against the 6.7\,yr period. This time, the variations appear much better phased: all \xmm\ data from 2004 lie close together, with the \xmm\ observation taken seven years later, in 2011; the two \xmm\ datasets from 2007 appear brighter, but at another phase (Fig. \ref{flux5}).

One might wonder whether these 2007 data are not corrupted in some way, as they are the only data in which the source is very bright. This scenario can be rejected for several reasons: (1) the two datasets were not taken on the same satellite orbit, rendering a technical problem unlikely, (2) the two datasets do not yield a peculiar brightening for \cygh\ or \cygd\ (this paper), or the other sources of the field \citep{rau11,naz12}, (3) {\it Suzaku} observations taken in 2007 confirm the brightening at that time of \cygc\ (Table \ref{fit5}, confirming \citealt{yos11}). Therefore, we analyzed the results further by considering that the X-ray emission could be modulated with the 6.7\,yr period. In addition to an increased flux, we found a stronger absorption at $\phi<0.2$ (Fig. \ref{flux5}). For a WWC, the flux and absorption variations are linked to the orbit of the stars: our findings can thus be related to orbital parameters.

Finally, one might wonder why the two \xmm\ data from 2007 yield such different results. Several scenarios are possible: either these short-term changes are due to inhomogeneities crossing the shock (as in $\eta$\,Car, \citealt{mof09}), or these variations suggest that part of the X-ray emission is linked to another WWC, in the binary or between the three stars and the fourth one. 

\subsection{Toward an orbital solution for the third star in \cygc?}
While the orbit of the eclipsing binary is now well known \citep{lin09}, the orbit of the third star around the binary system is not well constrained. \citet{ken10} proposed four solutions, which could explain the observed changes in radio emission, but the uncertainties on their parameters are high: an error of 4\% is associated with the period, while the error on the inclination is about 40$^{\circ}$, and eccentricity ranges from $\sim$0.1 to $\sim$0.7, time of periastron varies by 320 days, and the argument of periastron ranges from 23 to 352$^{\circ}$, depending on the chosen orbital solution. The derived X-ray properties can help further constrain the orbital parameters.

\begin{figure}
   \centering
   \includegraphics[width=9cm, bb=34 131 437 656, clip]{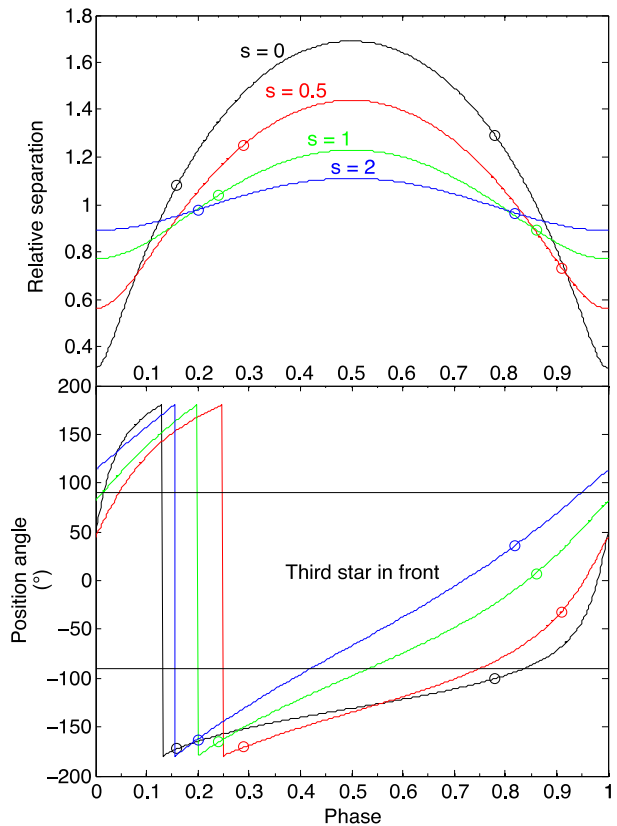}
      \caption{Evolution with phase of the relative distance between the contact binary system and the third star (top) and of their relative position (bottom) for the four orbital solutions proposed for \cygc\ by \citet{ken10}. The position angle is zero when the third star is in front of the binary, and 180$^{\circ}$ when the binary is in front of the third star. The circles indicate the position of the 2007 \xmm\ observation (left side) and of the 2011 \xmm\ observations (right side).}
         \label{reldist5}
   \end{figure}

Absorption variations are expected when the WWC zone and its associated X-ray emission are seen through different winds. The orbital solutions from \citet{ken10} are able to explain the larger absorption at $\phi<0.2$: at these phases, the contact binary system is in front (see lower panel of Fig. \ref{reldist5}), i.e., on the part of its orbit located toward the observer, so that the WWC zone is seen through its denser wind. For a long-period system, we also expect an increase in flux as the separation between the components decreases \citep{ste92}. The exact amount of change directly depends on the eccentricity of the system. The four solutions of \citet{ken10} have very different eccentricities, so that they predict very different variations in flux.

\clearpage
\begin{sidewaystable}
\caption[]{Results of the X-ray spectral fitting for \cygc. }
\label{fit5}
\centering
\begin{tabular}{ccccccccccccccccc }
\hline\hline
\multirow{3}{*}{Facility} & \multirow{3}{*}{Observation ID} &  \multirow{2}{*}{$n_{\text{H}}^{\text{add}}$} &  \multirow{2}{*}{\textit{norm}$_{1}$} &  \multirow{2}{*}{\textit{norm}$_{2}$} & \multicolumn{3}{c}{$F_{X}^{\text{obs}}$} &  \multicolumn{3}{c}{$F_{X}^{\text{unabs}}$} & \multirow{3}{*}{$\chi^{2}_{\nu}$ (d.o.f.)} \\
 &&&&&\multicolumn{3}{c}{(10$^{-12}$ erg cm$^{-2}$ s$^{-1}$)}&\multicolumn{3}{c}{(10$^{-12}$ erg cm$^{-2}$ s$^{-1}$)}&\\
 &&($10^{22}$ cm$^{-2}$)&($10^{-2}$ cm$^{-5}$)&($10^{-3}$ cm$^{-5}$)&Total&Soft&Hard&Total&Soft&Hard&\\
\hline
\multirow{7}{*}{\xmm}
 &0200450201&{0.51$_{-0.04}^{+0.04}$}& {9.50$_{-1.21}^{+1.26}$}& {6.19$_{-0.18}^{+0.18}$}&{2.64$\pm0.03$}& {1.41$\pm0.02$}& {1.23$\pm0.03$}& {15.58$\pm0.15$}& {14.04$\pm0.15$}& {1.57$\pm0.03$}&{1.36}\hspace{1.5mm}({323})\\
 &0200450301&{0.53$_{-0.04}^{+0.04}$}& {10.4$_{-1.3}^{+1.4}$}  & {6.43$_{-0.18}^{+0.18}$}&{2.72$\pm0.03$}& {1.44$\pm0.02$}& {1.28$\pm0.03$}& {15.40$\pm0.15$}& {13.80$\pm0.15$}& {1.60$\pm0.03$}& {1.71}\hspace{1.5mm}({327})\\
 &0200450401&{0.49$_{-0.03}^{+0.03}$}& {9.51$_{-0.98}^{+0.98}$}& {6.02$_{-0.15}^{+0.15}$}&{2.61$\pm0.02$}& {1.41$\pm0.01$}& {1.20$\pm0.02$}& {16.05$\pm0.13$}& {14.54$\pm0.13$}& {1.51$\pm0.03$}&{1.29}\hspace{1.5mm}({363})\\
 &0200450501&{0.50$_{-0.04}^{+0.04}$}& {9.42$_{-1.35}^{+1.42}$}& {6.37$_{-0.20}^{+0.20}$}&{2.70$\pm0.03$}& {1.43$\pm0.02$}& {1.27$\pm0.03$}& {15.92$\pm0.16$}& {14.32$\pm0.17$}& {1.59$\pm0.04$}&{1.24}\hspace{1.5mm}({299})\\
 &0505110301&{0.69$_{-0.05}^{+0.05}$}& {14.2$_{-2.3}^{+2.5}$}  & {11.1$_{-0.3}^{+0.3}$}&{3.95$\pm0.04$}& {1.84$\pm0.02$}& {2.11$\pm0.05$}& {14.87$\pm0.17$}& {12.22$\pm0.16$}& {2.64$\pm0.06$}&{1.27}\hspace{1.5mm}({302})\\
 &0505110401&{0.70$_{-0.04}^{+0.04}$}& {19.1$_{-2.3}^{+2.4}$}  & {14.1$_{-0.3}^{+0.3}$}&{5.05$\pm0.04$}& {2.36$\pm0.02$}& {2.69$\pm0.05$}& {19.11$\pm0.17$}& {15.71$\pm0.16$}& {3.36$\pm0.06$}& {1.40}\hspace{1.5mm}({378})\\
 &0677980601&{0.52$_{-0.04}^{+0.04}$}& {10.9$_{-1.3}^{+1.4}$}  & {6.83$_{-0.20}^{+0.20}$}&{2.92$\pm0.03$}& {1.56$\pm0.02$}& {1.36$\pm0.03$}& {17.17$\pm0.16$}& {15.47$\pm0.17$}& {1.71$\pm0.04$}&{1.81}\hspace{1.5mm}({247})\\
\hline
\multirow{7}{*}{\sw}
&00031904001    &{0.60$_{-0.10}^{+0.11}$}& {10.1$_{-1.4}^{+1.5}$}  &&{2.64$\pm0.14$}& {1.32$\pm0.08$}& {1.33$\pm0.14$}& {12.5$\pm0.6$}& {10.8$\pm0.6$ }& {1.67$\pm0.18$}&{0.71}\hspace{1.5mm}({36})\\
&00031904002    &{0.44$_{-0.11}^{+0.12}$}& {6.69$_{-1.11}^{+1.18}$}&&{2.00$\pm0.12$}& {1.09$\pm0.07$}& {0.91$\pm0.13$}& {13.8$\pm0.8$}& {12.6$\pm0.8$} & {1.15$\pm0.17$}&{1.32}\hspace{1.5mm}({30})\\
&00031904003    &{0.32$_{-0.10}^{+0.11}$}& {7.91$_{-1.23}^{+1.34}$}&&{2.65$\pm0.14$}& {1.54$\pm0.10$}& {1.11$\pm0.13$}& {25.9$\pm1.4$}& {24.5$\pm1.5$} & {1.40$\pm0.16$}&{0.95}\hspace{1.5mm}({33})\\
&00031904004    &{0.42$_{-0.12}^{+0.13}$}& {7.57$_{-1.29}^{+1.42}$}&&{2.31$\pm0.13$}& {1.28$\pm0.08$}& {1.04$\pm0.13$}& {17.0$\pm0.9$}& {15.7$\pm1.0$} & {1.31$\pm0.17$}&{0.73}\hspace{1.5mm}({31})\\
&00031904005    &{0.60$_{-0.13}^{+0.14}$}& {10.4$_{-1.6}^{+1.8}$}  &&{2.74$\pm0.15$}& {1.37$\pm0.09$}& {1.38$\pm0.13$}& {13.0$\pm0.7$}& {11.3$\pm0.7$ }& {1.73$\pm0.17$}&{1.02}\hspace{1.5mm}({34})\\
&00031904001--5 &{0.46$_{-0.05}^{+0.05}$}& {8.39$_{-0.58}^{+0.60}$}&&{2.46$\pm0.06$}& {1.32$\pm0.04$}& {1.14$\pm0.06$}& {16.0$\pm0.4$}& {14.6$\pm0.4$} & {1.43$\pm0.07$}&{1.13}\hspace{1.5mm}({138})\\
&00032767001$+$2&{0.83$_{-0.11}^{+0.12}$}& {13.9$_{-1.7}^{+1.8}$}  &&{3.14$\pm0.13$}& {1.38$\pm0.07$}& {1.75$\pm0.13$}& {10.1$\pm0.4$}& {7.94$\pm0.41$}& {2.20$\pm0.16$}&{1.31}\hspace{1.5mm}({51})\\
\hline
\multirow{6}{*}{\it ROSAT}
 &200109        &{0.45$_{-0.14}^{+0.15}$}& {7.90$_{-1.89}^{+2.36}$}&&{/}& {1.27$\pm0.07$}& {/}& {/}& {14.3$\pm0.8$}& {/}&{1.01}\hspace{1.5mm}({33})\\
 &900314        &{0.59$_{-0.06}^{+0.06}$}& {10.5$_{-1.0}^{+1.1}$}&&{/}  & {1.40$\pm0.03$}& {/}& {/}& {11.8$\pm0.3$}& {/}& {1.08}\hspace{1.5mm}({125})\\
 &{900314} : 1/4&{0.81$_{-0.20}^{+0.24}$}& {14.0$_{-3.9}^{+5.5}$}&&{/}  & {1.40$\pm0.10$}& {/}& {/}& {8.19$\pm0.60$}& {/}& {1.22}\hspace{1.5mm}({20})\\
 &{900314} : 2/4&{0.63$_{-0.14}^{+0.15}$}& {10.6$_{-2.3}^{+2.8}$}&& {/} & {1.35$\pm0.07$}&{/}& {/}&  {10.6$\pm0.6$}& {/}& {0.88}\hspace{1.5mm}({35})\\
 &{900314} : 3/4&{0.60$_{-0.08}^{+0.08}$}& {11.4$_{-1.5}^{+1.7}$}&& {/} & {1.50$\pm0.05$}& {/}& {/}& {12.4$\pm0.4$}& {/}& {1.27}\hspace{1.5mm}({68})\\
 &{900314} : 4/4&{0.48$_{-0.10}^{+0.11}$}& {8.61$_{-1.54}^{+1.80}$}&&{/}& {1.33$\pm0.06$}& {/}& {/}& {14.1$\pm0.6$}& {/}& {1.00}\hspace{1.5mm}({54})\\
\hline
{\it Suzaku} & 402030010 & {0.57$_{-0.02}^{+0.02}$}& {13.6$_{-0.3}^{+0.3}$}&&{3.65$\pm0.03$}& {1.85$\pm0.02$}& {1.81$\pm0.03$}& {18.4$\pm0.2$}& {16.1$\pm0.2$}& {2.27$\pm0.04$}& {1.31}\hspace{1.5mm}({825})\\
\hline
\end{tabular}
\tablefoot{The fitted model has the form $wabs*phabs*(apec+apec)$, with the first absorption fixed to $1.14\times10^{22}$\,cm$^{-2}$, the temperatures fixed to 0.22 and 1.19\,keV, and the ratio of normalization factors $norm_1/norm_2$ fixed to 14.93 for the \sw, {\it Suzaku}, and {\it ROSAT} data. Error bars and energy bands are the same as in Table \ref{fit8}.}
\clearpage
\end{sidewaystable}

\clearpage

The two solutions with low eccentricities ($s=1$ or 2 in \citealt{ken10}) imply limited changes in separation (see upper panel of Fig. \ref{reldist5}), incompatible with the large observed variation in flux. The intermediate solution ($s=0.5$ in \citealt{ken10}) would lead to a change in flux opposite to what is seen in the data (i.e., the source being much brighter in 2004 and 2011 than in 2007), so that this solution can be discarded too. 

The solution favored by \citet{ken10}, that with $s=0$, is thus also our favorite. However, its ephemeris is certainly not perfect. Indeed, we would expect the largest flux for the second {\it ROSAT} dataset (ObsID 900314), since it was taken close to the expected periastron passage in the 6.7\,yr period (Fig. \ref{flux5}). Such a flux is not observed, but {\it ROSAT} is not sensitive to hard X-rays, so that we requested a new \sw\ observation, which confirmed the {\it ROSAT} results. These new data show only a moderate increase in hard flux (about 35\%) relative to observations at $\phi\sim0.8$, while a change by a factor of 2.6 is expected. Moreover, the variation between the \xmm\ data taken in 2007 (Revs 1353 and 1355) and the observations at $\phi\sim0.8$ amounts to 85\%, whereas the solution by \citet{ken10} suggests 20\%. Therefore, if a high eccentricity is favored for the orbit of the third star, a revised ephemeris is certainly needed.

\begin{figure*}
   \centering
   \includegraphics[width=8.2cm,height=6cm,bb=19 164 544 621,clip]{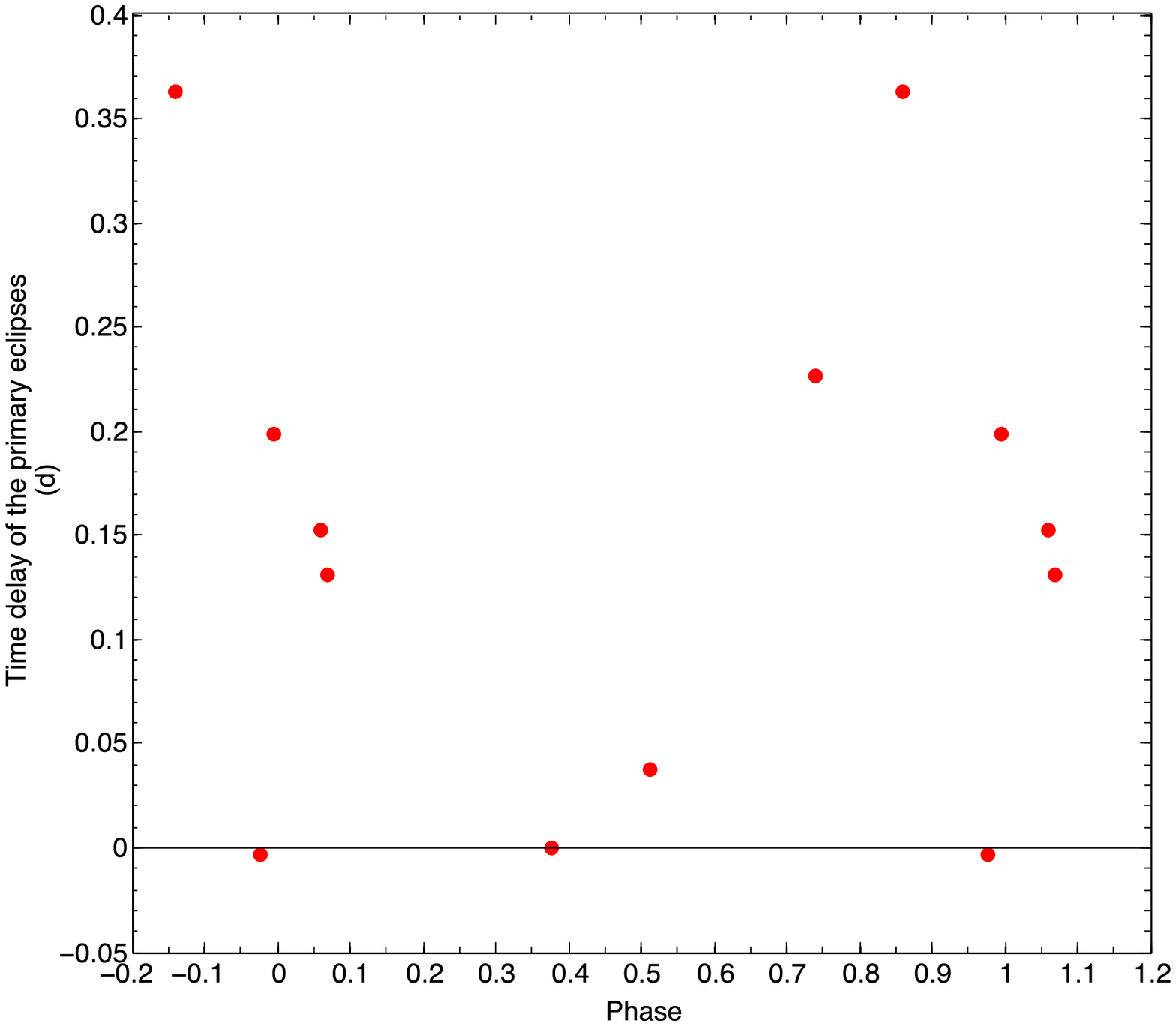}\hspace{0.3cm}
   \includegraphics[width=8.2cm,height=6cm,bb=0 170 551 517,clip]{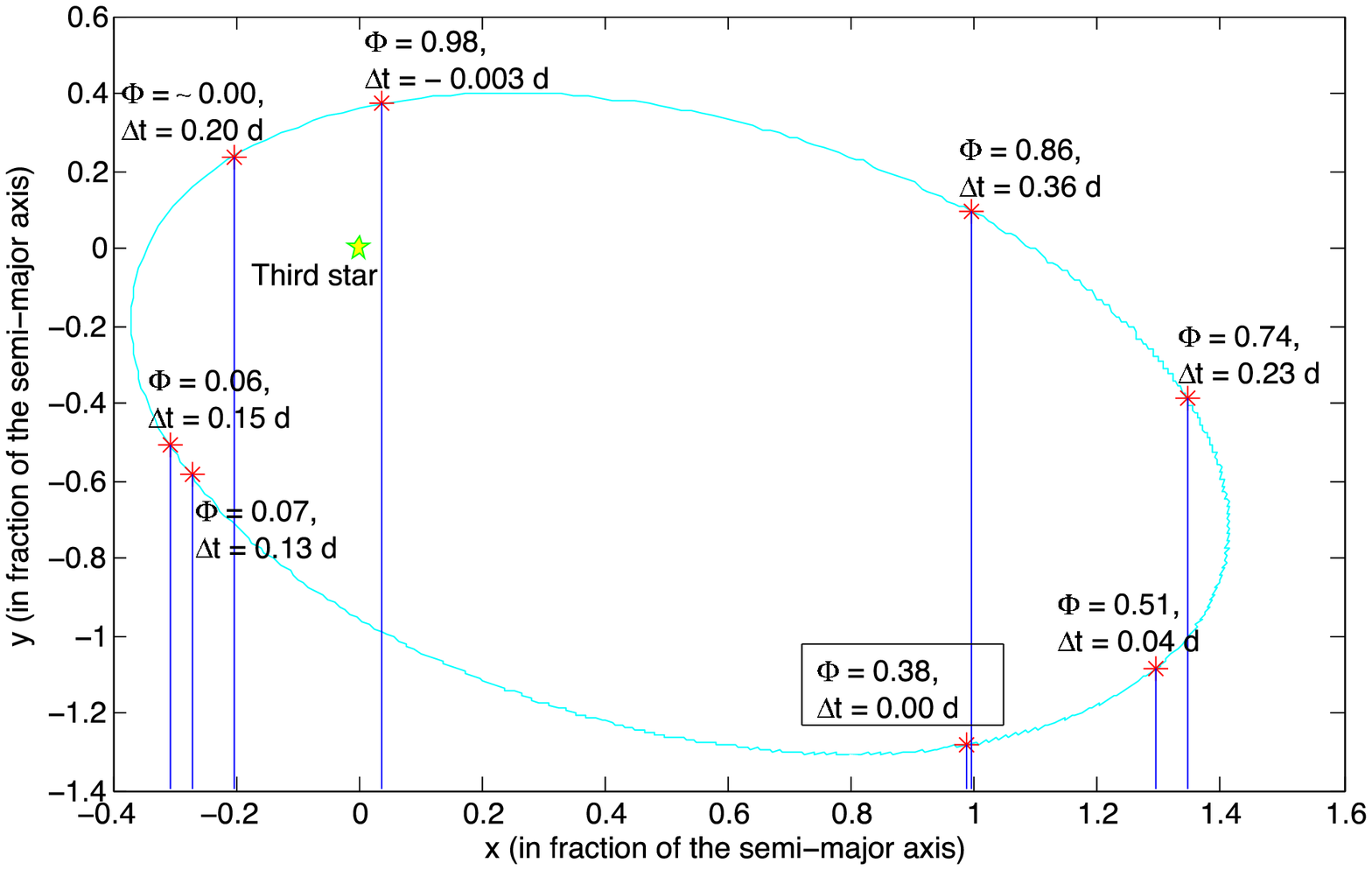}
      \caption{{\it Left:} evolution with phase of the difference in time between expected and observed eclipses (primary minima) of \cygc. {\it Right:} orbit of the binary (in the favored solution of \citealt{ken10}) along with these temporal differences. The observer is toward the bottom.}
         \label{delay5}
   \end{figure*}

To complement the X-ray monitoring, we have made the first steps in this direction by considering the reflex motion of the binary due to the presence of the third star. For example, the eclipses provide a clock that can be used to find temporal delays produced by light travel time effects linked to orbital motion. To this aim, we searched for times of primary minima in the literature (Table \ref{eclipses}). Choosing one of these dates as (arbitrary) reference, we calculated the time difference between the other observations and the reference (Fig. \ref{delay5}). We find that the delay increases as the binary recedes, as expected (Fig. \ref{delay5}). Linder et al. (2009) adopted an alternative working hypothesis, where a change of the orbital period due to mass loss leads to quadratic ephemerides. To determine these quadratic ephemerides, Linder et al. (2009) used the times of primary eclipse quoted by Hall (1974), as well as their own data and the result of H\"ubscher \& Walter (2007). 
However, the fit with the quadratic ephemerides inconsistent with the three data points of Sazonov (1961) and the measurement of H\"ubscher \& Walter (2007) that led to large residuals. With our new interpretation, the data of Sazonov (1961) and H\"ubscher \& Walter (2007) follow the trend expected for a time delay in a triple system. However, a large deviation is now found for the point of Miczaika (1953). This is surprising as the observations of  the latter author were taken roughly at the same time as the third epoch quoted by Sazonov (1961). Unfortunately, the phase coverage of the light curve of Miczaika (1953) is quite limited and the time of primary minimum is simply quoted as the time of minimum light in the data collected by this author. Generally speaking, the uncertainties on the determination of the times of primary minimum from most of the archival data are difficult to estimate. Therefore, a new, precise monitoring of the system is required to further improve the orbital solution.

\begin{table}
\caption{Time delays between dates of primary minima of the inner binary system of \cygc\ found in the literature and a reference date.}
\label{eclipses}
\centering
\begin{tabular}{ccccccc}
\hline\hline
\multirow{2}{*}{Source} & HJD$_{0}$  & Phase (for & {$\Delta t$} \\
 & $- 2400000$& $P$=6.7\,yrs) & (d) \\
\hline
\citet{saz61} & 28749.154 & 0.74 & 0.23 \\
\citet{saz61} & 29553.985 & 0.07 & 0.13  \\
\citet{wil51} & 32747.167 & 0.38 & 0.00  (Ref.) \\
\citet{mic53} & 34218.463 & 0.98 & $-3\times 10^{-3}$  \\
\citet{saz61} & 34264.849 & 0.00 & 0.20\\
\citet{hal74} & 40413.796 & 0.51 & 0.04 \\
\citet{lin09} & 51049.702 & 0.86 & 0.36 \\
\citet{hub07} & 53985.493 & 0.06 & 0.15 \\
 \hline
\end{tabular}
\end{table}

\citet{ken10} had already compared the observed radial velocities of the binary components in \cygc\ with the predicted radial velocities of a mean orbital solution. They showed that their favorite solution is compatible with the observed residuals, but the large error bars (7 -- 9\,km\,s$^{-1}$ to be compared with their maximum velocity difference of 22.5\,km\,s$^{-1}$) and incomplete phase coverage (periastron passage not covered) could not lead to a significant statistical test. Following the $s=0$ solution of \citet{ken10}, a more significant variation could be expected around periastron (see their Fig.\ 11) and the first semester of 2013 is close to the expected periastron passage in their favorite solution. 

We have thus obtained optical spectra over six consecutive nights in June 2013. Our observing campaign corresponds to orbital phase 0.07 of the 6.7\,yr cycle (according to the $s=0$ ephemeris of \citealt{ken10}). The spectra were taken with the Aur\'elie spectrograph \citep{gil94} at the 1.52\,m telescope of the OHP. We used a 600 lines\,mm$^{-1}$ grating blazed at 5000\,\AA\ and covered the wavelength domain from 4448 to 4886\,\AA\ at a resolving power of $\simeq 10000$. The detector was an EEV 42-20 CCD with $2048 \times 1024$ pixels. The data reduction was done in the standard way (see \citealt{bd60}), using the MIDAS software provided by ESO.

\begin{figure}
\begin{center}
\includegraphics[width=8cm]{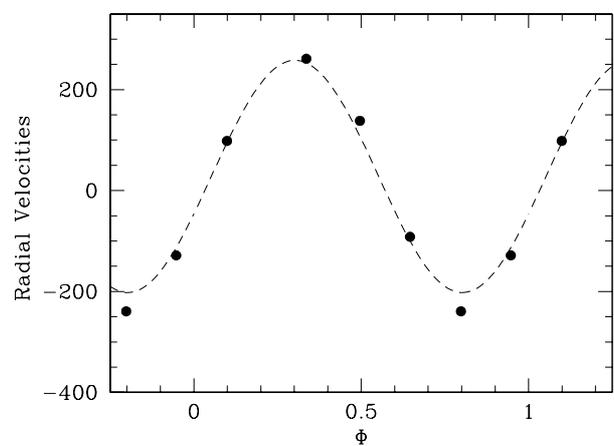}
\end{center}
\caption{Radial velocities of the He\,{\sc ii} $\lambda$\,4686 emission line as measured on our June 2013 optical spectra of Cyg\,OB2 \#5. The S-wave relation from \citet{Rauw} is shown by the dashed line. No systematic shift is found between the data and the curve, while the $s = 0$ orbital solution of \citet{ken10} predicts a shift of about 35\,km\,s$^{-1}$ toward negative values.\label{HeII4686}}
\end{figure}

Assessing the velocities of the binary components is not easy, however: different lines in the spectra of components A and B often yield different systemic velocities and some of the secondary's absorption lines (H\,{\sc i}, He\,{\sc i}) display P-Cygni profiles over some parts of the orbital cycle \citep{Rauw}. A reliable indicator of a change in systemic velocity could, however, be the velocity of the peak of the He\,{\sc ii} $\lambda$\,4686 emission line. Indeed, this line undergoes strong profile variations over the 6.6\,d cycle, but the radial velocities of its peak were found to describe a very stable sine-wave variation following the primary's motion (see Figs.\ 3 and 4 of \citealt{Rauw}). Fig.\,\ref{HeII4686} illustrates the radial velocities that we measured on the June 2013 data, along with the best-fit S-wave curve from \citet{Rauw}, i.e., not a new fit\footnote{Note that the scatter around the S-wave curve is comparable to that found in the old data.}! In fact, the shift between the 
new and old 
data amounts to $(0.5 \pm 24.7)$\,km\,s$^{-1}$, indicating that there is no net shift in the systemic velocity of the June 2013 data with respect to the older data of \citet{Rauw}, despite their different phases in the 6.7\,yr cycle. If the $s=0$ solution of \cite{ken10} was correct, one would rather expect a systemic velocity shifted by about 35\,km\,s$^{-1}$ toward the negative values. As the X-ray data suggested, the $s=0$ solution favored by \citet{ken10} is far from perfect, underlining the need to collect spectroscopic data over the full 6.7\,yr radio cycle, to establish the orbital solution of the third component.  

\subsection{Revised light curve of \cygc}
In principle, the analysis of the light curve of an eclipsing binary yields the absolute dimensions of its components. These numbers can then be used to infer the distance of the binary system. This approach was adopted by \citet{lin09} to derive a distance of d = $(925 \pm 0.25)$\,pc (DM = $9.83 \pm 0.06$) for Cyg\,OB2 \#5. This number is significantly lower than the distance estimates of Cyg\,OB2 found in the literature (DM = $11.2 \pm 0.1$, \citealt{massey}, \citealt{Kiminki}; DM = $10.4$, \citealt{hanson}; d = $(1.40 \pm 0.08)$\,kpc, \citealt{Rygl}). Recently, \citet{Dzib} determined a distance of d = $1.65^{+0.96}_{-0.44}$\,kpc for Cyg\,OB \#5. This distance estimate was obtained from the trigonometric parallax of the most compact radio source in the system. 

\begin{figure}
\centering
\includegraphics[width=8cm,bb=1 1 558 443, clip]{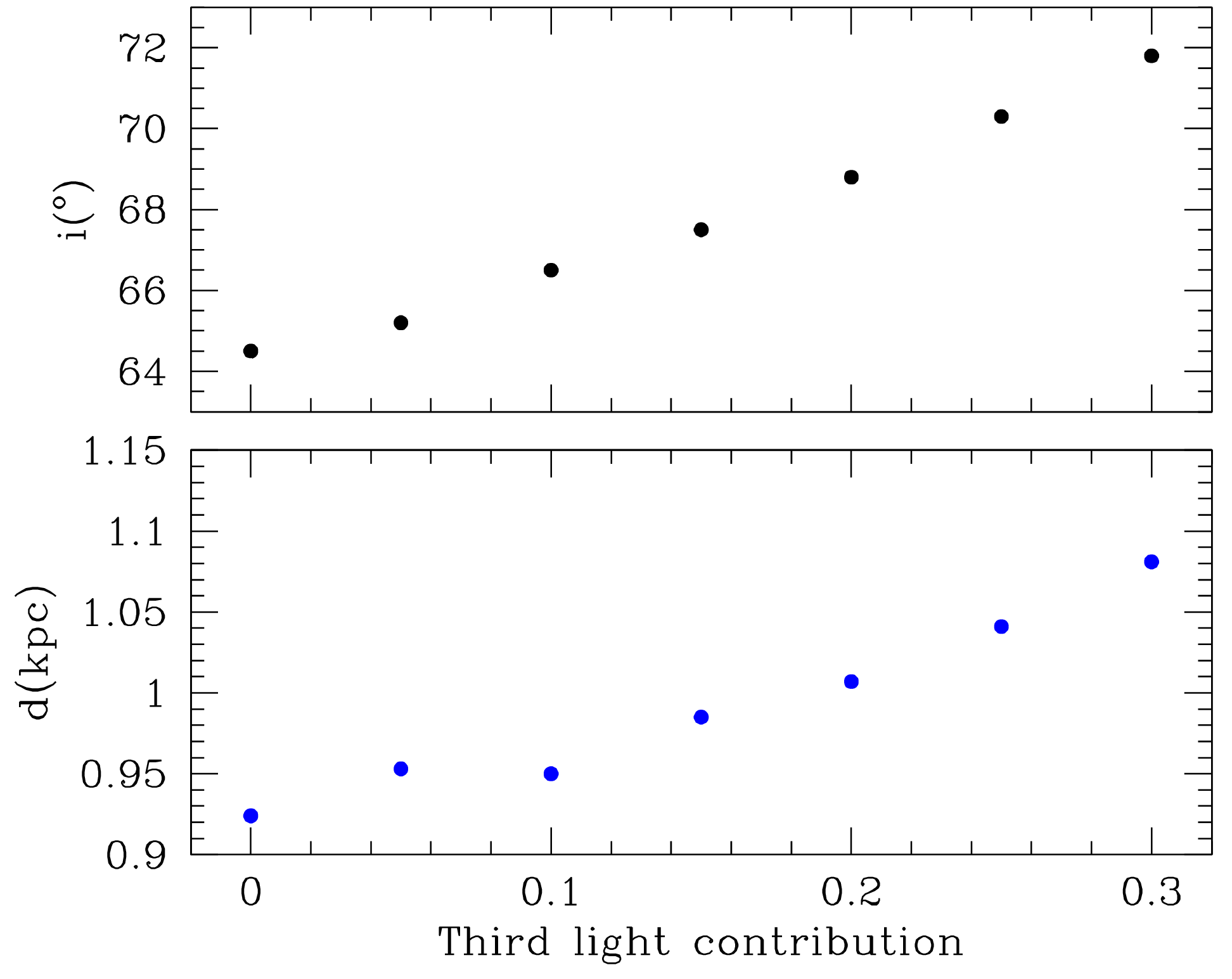}
\caption{{\it Top:} the best-fit orbital inclination as a function of the third light contribution. Typical uncertainties on $i$ are $1.5^{\circ}$. {\it Bottom:} distance of Cyg\,OB2 \#5 inferred from the light curve analysis as a function of the third light contribution.\label{thirdlight}}
\end{figure}

However, the analysis of \citet{lin09} could be biased by the presence of a third light. Indeed, Cyg\,OB2 \#5 was subsequently found to most likely consist of four stars \citep{ken10}. The astrometric companion (star D in \citealt{ken10}) is too faint to play a role in the combined light \citep{con97}. The properties of the third component (star C) are poorly constrained, although it seems likely that this could be a luminous late O-type star \citep{ken10}. 

The depths of the primary and secondary eclipses (about 0.35 and 0.25\,mag, respectively) leave significant room for a third light contribution. Indeed, assuming that both eclipses would be total eclipses, we find an upper limit on the third light of 52\% of the total light. Yet, if star C were that bright, it should dominate the spectrum of Cyg\,OB2 \#5. Although we cannot rule out the presence of a third system of lines, the observed spectra of Cyg\,OB2 \#5 are dominated by features that exhibit the signature of the 6.6\,d orbital cycle \citep{Rauw}. Therefore, the contribution of star C is very likely below about 30\%. To the zeroth order, the distance is expected to vary as $\sqrt{1 + l_C/(l_A+l_B)}$ where $l_A$, $l_B$ and $l_C$ are the fluxes of components A, B (the primary and secondary of the eclipsing binary), and C, respectively, in the waveband under consideration. However, the presence of a third light in the light curve also affects the best-fit parameters inferred for components A and B, and a 
full 
analysis is thus required to evaluate the impact of the third light on the distance estimate. 
 
We have thus repeated the analysis of the light curve presented by \citet{lin09} accounting this time for the contribution of a nonzero third light. The analysis was performed with the {\tt NIGHTFALL} code (version 1.70) developed and maintained by Wichmann, Kuster, and Risse\footnote{\tiny http://www.hs.uni-hamburg.de/DE/Ins/per/Wichmann/Nightfall.html}. The assumptions were the same as in the work of \citet{lin09}: both stars of the eclipsing binary are in a contact or overcontact configuration with identical Roche-lobe filling factors, the primary star has an effective temperature of 36000\,K, the secondary star features an extended hot spot on the side facing the primary (latitude and longitude fixed at $0^{\circ}$ and $-15^{\circ}$, respectively). We varied the third light contribution $l_C/(l_A+l_B+l_C)$ between 0 and 30\% by steps of 5\%. For each assumed third light contribution, we obtained the best fit to the observed continuum band light curves. 

As one could expect, the best-fit orbital inclination increases monotonically with increasing contribution of the third light (see Fig.\,\ref{thirdlight}). The best-fit Roche-lobe filling factor also increases slightly from 1.00 to 1.03 when the third light contribution varies from 0 to 30\%. The best-fit parameters of the binary components thus change (this is especially the case for the secondary star), but their overall properties remain in qualitative agreement with the description of \citet{lin09}. The lower panel of Fig.\,\ref{thirdlight} indicates the dependence of the distance with the third light contribution. This figure shows that even for a 30\% contribution, the distance `only' increases to 1.1\,kpc. We thus conclude that third light alone cannot bring the distance of the eclipsing binary into agreement with other distance estimates. 

If Cyg\,OB2 \#5 is indeed a member of Cyg\,OB2, which seems rather likely, other factors must play a role. For instance, an obvious factor could be a higher primary star temperature. Indeed, the temperature estimate could be biased by contamination of the spectrum by component C. Assessing an unbiased temperature estimate requires disentangling the spectra of components A, B, and C. This is currently not possible though, as we are lacking sufficient coverage of the 6.7\,yr cycle. Another factor could be uncertainties on the reddening. Here, we have adopted $A_V = 6.37$ (see discussion in \citealt{lin09}), which corresponds to $R_V = 3.27$. Reducing $R_V$ to $3.1$ would increase the distance modulus by 0.33\,mag, and thus the distance, by 16\%.

\section{The object \cygd }
\label{section_cygob212}
The star \cygd\ is one of the brightest stars in the Galaxy (e.g.,\ \citealt{massey}). Though it is a very luminous hot star, its classification as a Luminous Blue Variable (LBV) remains debated. Indeed, it lacks some of the typical LBV characteristics \citep{cla11}. Its bright (\loglxlb$=-6.1$, \citealt{rau11}) X-ray emission is also unusual for such objects, considering its wind properties and its isolated nature \citep{naz12lbv}. 

\citet{rau11} analyzed the first six \xmm\ observations, finding high temperatures and some variability. These properties are reminiscent of WWCs or of magnetically confined winds, though no sign of binarity or magnetic field was detected up to now for \cygd.

\begin{figure}
   \centering
 
     \includegraphics[scale=0.3]{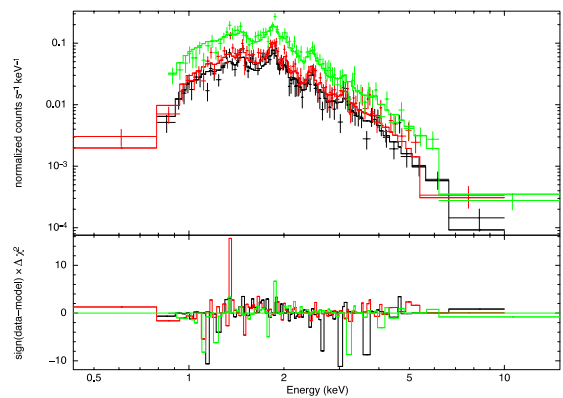}
 \caption{\cygd\ acquired with the MOS1 (black), MOS2 (red), and pn (green) detectors in May 2007 (Rev. 1355) along with the best-fit model and its residuals.}
         \label{cyg12spec}
   \end{figure}

Our dataset improves the temporal coverage of X-ray studies, leading to a revision of the known properties of \cygd. We used an equivalent H absorbing column of $1.97 \times 10^{22}$\,cm$^{-2}$ (corresponding to $E(B-V)=3.40$, \citealt{van01}) and first performed spectral fitting of the \xmm\ data only, using two thermal components. The best-fit values of the additional absorption are zero or compatible with zero within the uncertainties. Furthermore, the best-fit temperatures are high: 0.86 and 2.11\,keV. They appear slightly lower in the last \xmm\ dataset, but the difference does not appear significant if we take the error bars into account. Therefore, we fixed the temperatures, discarded the additional absorption, and fitted all \xmm\ spectra (Table \ref{fit12}, Fig. \ref{cyg12spec}) again. The ratio between the two normalization factors remains similar, within the uncertainties, amongst the observations, so that we further fixed it to $norm_1/norm_2\sim3.26$ for fitting the \sw, {\it ROSAT}, 
and {\it Suzaku} spectra (Table \ref{fit12}).

\begin{sidewaystable}
\caption{Results of the X-ray spectral fitting for \cygd.}
\label{fit12}
\centering
\begin{tabular}{ccccccccccccccccc }

\hline\hline
\multirow{3}{*}{Facility} & \multirow{3}{*}{Observation ID} &  \multirow{2}{*}{\textit{norm}$_{1}$} &  \multirow{2}{*}{\textit{norm}$_{2}$} & \multicolumn{3}{c}{$F_{X}^{\text{obs}}$} &  \multicolumn{3}{c}{$F_{X}^{\text{unabs}}$} & \multirow{3}{*}{$\chi^{2}_{\nu}$ (d.o.f.)} \\

 &&&&\multicolumn{3}{c}{(10$^{-12}$ erg cm$^{-2}$ s$^{-1}$)}&\multicolumn{3}{c}{(10$^{-12}$ erg cm$^{-2}$ s$^{-1}$)}&\\

 &&($10^{-3}$ cm$^{-5}$)&($10^{-3}$ cm$^{-5}$)&Total&Soft&Hard&Total&Soft&Hard&\\
\hline
\multirow{7}{*}{\xmm}
 &0200450201&{7.24$_{-0.34}^{+0.34}$}& {2.41$_{-0.20}^{+0.20}$}&{2.52$\pm0.03$}& {0.81$\pm0.01$}& {1.71$\pm0.04$}& {25.1$\pm0.3$}& {22.7$\pm0.4$}& {2.42$\pm0.05$}& {1.23}\hspace{1.5mm}({218})\\
 &0200450301&{7.13$_{-0.25}^{+0.25}$}& {2.49$_{-0.15}^{+0.15}$}&{2.54$\pm0.03$}& {0.82$\pm0.01$}& {1.72$\pm0.03$}& {24.9$\pm0.2$}& {22.4$\pm0.3$}& {2.43$\pm0.04$}& {1.14}\hspace{1.5mm}({257})\\
 &0200450401&{8.32$_{-0.25}^{+0.25}$}& {2.61$_{-0.14}^{+0.14}$}&{2.82$\pm0.03$}& {0.94$\pm0.01$}& {1.88$\pm0.02$}& {28.6$\pm0.3$}& {25.9$\pm0.3$}& {2.67$\pm0.03$}& {1.36}\hspace{1.5mm}({266})\\
 &0200450501&{8.80$_{-0.32}^{+0.32}$}& {1.35$_{-0.17}^{+0.17}$}&{2.27$\pm0.03$}& {0.88$\pm0.01$}& {1.39$\pm0.03$}& {28.2$\pm0.4$}& {26.2$\pm0.4$}& {2.04$\pm0.05$}& {1.07}\hspace{1.5mm}({207})\\
 &0505110301&{5.11$_{-0.32}^{+0.32}$}& {1.46$_{-0.18}^{+0.18}$}&{1.66$\pm0.03$}& {0.56$\pm0.01$}& {1.10$\pm0.03$}& {17.4$\pm0.3$}& {15.8$\pm0.4$}& {1.56$\pm0.05$}& {1.29}\hspace{1.5mm}({206})\\
 &0505110401&{5.35$_{-0.27}^{+0.27}$}& {1.67$_{-0.16}^{+0.16}$}&{1.81$\pm0.03$}& {0.60$\pm0.01$}& {1.21$\pm0.03$}& {18.4$\pm0.3$}& {16.7$\pm0.3$}& {1.71$\pm0.05$}& {1.18}\hspace{1.5mm}({243})\\
 &0677980601&{5.10$_{-0.25}^{+0.25}$}& {1.35$_{-0.14}^{+0.14}$}&{1.60$\pm0.03$}& {0.56$\pm0.01$}& {1.04$\pm0.03$}& {17.2$\pm0.3$}& {15.7$\pm0.2$}& {1.48$\pm0.04$}& {1.13}\hspace{1.5mm}({110})\\
\hline
\multirow{2}{*}{\sw}
 &00031904001--5  &{6.04$_{-0.30}^{+0.30}$} & {} &{ 2.03$\pm0.06$}& { 0.67$\pm0.02$}& { 1.36$\pm0.06$}& { 20.7$\pm0.6$}& { 18.8$\pm0.7$}& { 1.94$\pm0.09$} & { 0.97}\hspace{1.5mm}({105})\\
 &00032767001$+$2 &{4.38$_{-0.51}^{+0.51}$} & {} &{ 1.47$\pm0.10$}& { 0.48$\pm0.04$}& { 0.99$\pm0.11$}& { 15.0$\pm1.0$}& { 13.6$\pm1.2$}& { 1.41$\pm0.16$}& { 1.23}\hspace{1.5mm}({23})\\
\hline
\multirow{2}{*}{\it ROSAT}
 & { 200109} & { 6.88$_{-0.95}^{+0.95}$} & {} & { /} & { 0.76$\pm0.06$} & { /} & { /} & { 21.4$\pm1.7$} & { /} & { 1.25}\hspace{1.5mm}({17})\\
 & { 900314} & { 7.33$_{-0.41}^{+0.41}$} & {} & { /} & { 0.81$\pm0.03$} & { /} & { /} & { 22.8$\pm0.8$} & { /} & { 0.87}\hspace{1.5mm}({78})\\
\hline
{\it Suzaku} & { 402030010} & { 7.78$_{-0.14}^{+0.14}$} & {} & { 2.61$\pm0.03$} & { 0.86$\pm0.01$} & { 1.75$\pm0.0$3} & { 26.7$\pm0.3$} & { 24.2$\pm0.3$} & { 2.50$\pm0.04$} & { 1.07}\hspace{1.5mm}({728})\\ 
\hline
\end{tabular}
\tablefoot{The fitted model has the form $wabs*(apec+apec)$, with the absorption fixed to $1.97\times10^{22}$\,cm$^{-2}$, the temperatures fixed to 0.86 and 2.11\,keV, and the ratio of normalization factors $norm_1/norm_2$ fixed to 3.26 for the \sw, {\it Suzaku}, and {\it ROSAT} data. Error bars and energy bands are the same as in Table \ref{fit8}. The MOS2 detector data of \cygd\ from three observations of 2004 (Revs. 0901, 0906, and 0911) were not taken into account as small differences between these data and the corresponding MOS1 and pn detectors data in a small zone around 1.3 keV were noticed.}
\end{sidewaystable}

   \begin{figure}
   \centering
   \includegraphics[width=8cm,bb=13 10 534 407,clip]{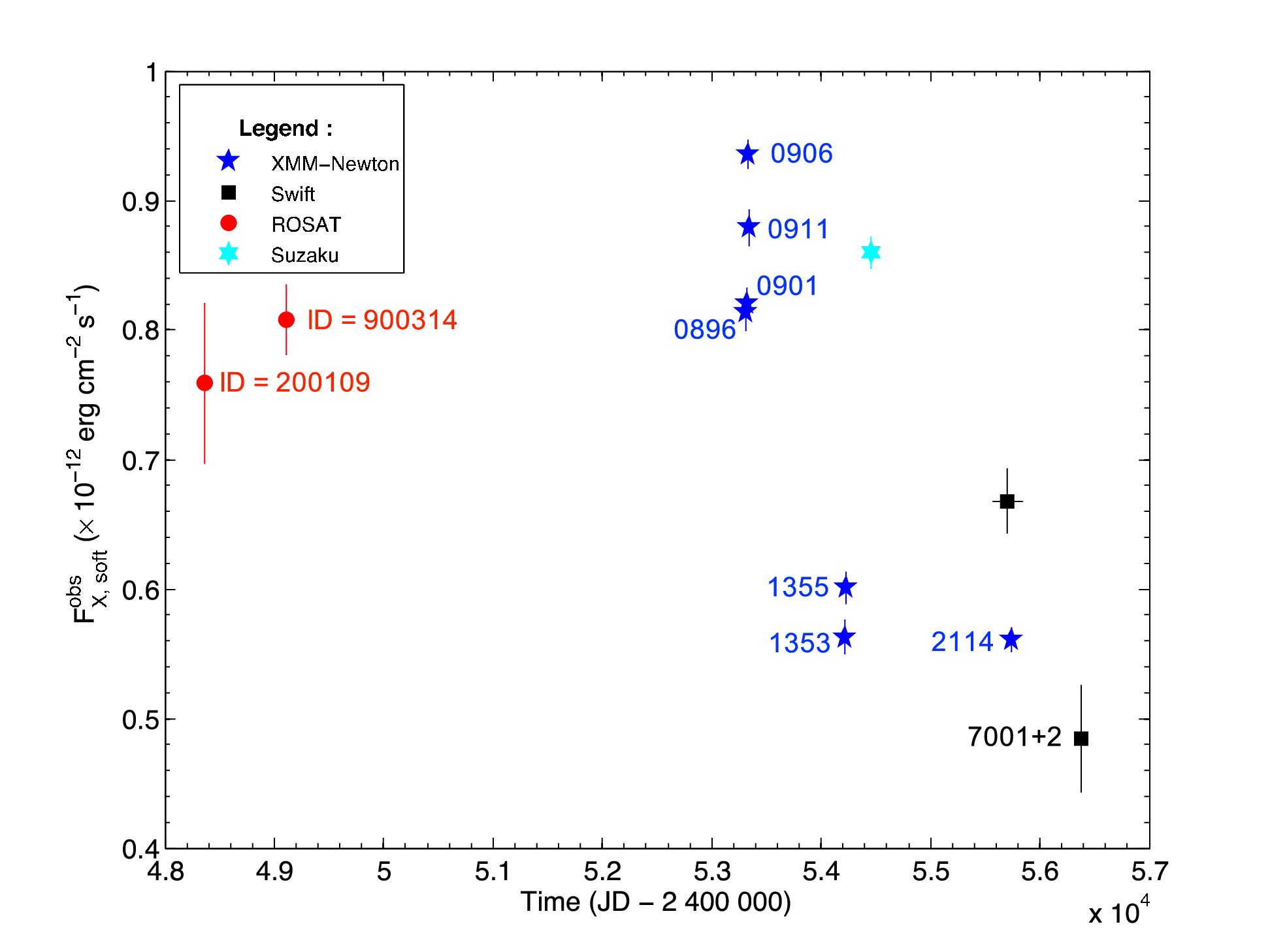}
   \includegraphics[width=8cm,bb=13 10 534 407,clip]{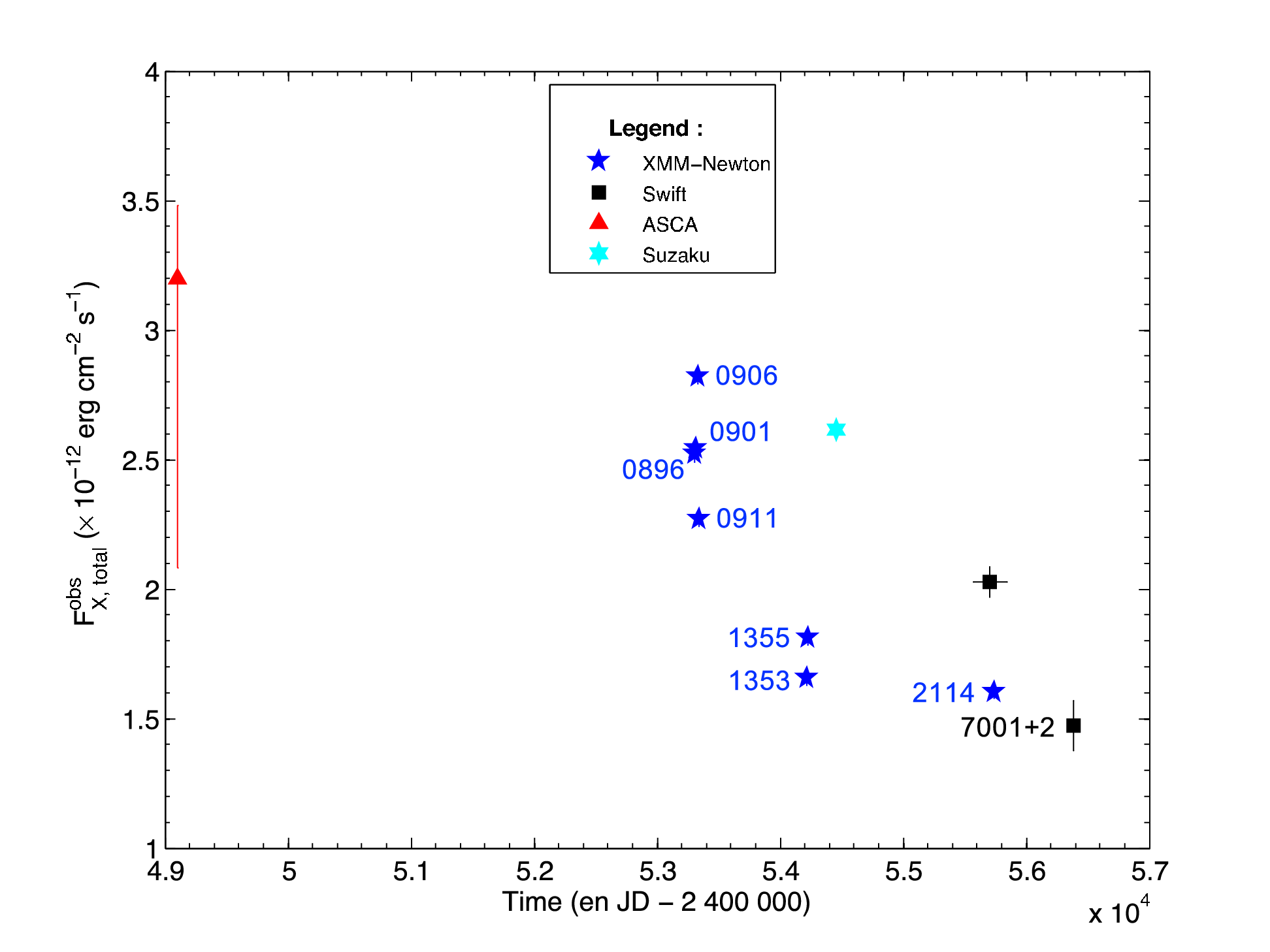}
      \caption{Evolution with time of the soft and total X-ray fluxes of \cygd. }
         \label{cyg12}
   \end{figure}

While the short-term variations reported by \citet{rau11} are confirmed, a long-term trend with larger amplitude is now detected (Fig. \ref{cyg12}). The observed X-ray flux decreases by 40\% between 2004 and 2011 in the \xmm\ data, and this decrease is confirmed by \sw\ and {\it Suzaku} data. Such a decrease can occur when the shocks linked to an eruption (an expected phenomenon for LBVs) settle down. However, no eruption was reported in the literature for \cygd\ in previous years, although the star does not appear to have been monitored so an eruption might have been missed. Another possibility to explain this decrease exists: a variation in the flux of a WWC as the two stars of an eccentric binary system recede from each other after periastron passage. In this case, \cygd\ would be a long-period binary, and its changes should be recurrent. Older {\it ROSAT} (Table \ref{fit12}) and the Advanced Satellite for Cosmology and Astrophysics ({\it ASCA}) \citep{yos11} observations taken in the 1990s 
indicate a brightening of \cygd, consistent with a WWC 
observed 
before periastron passage. The binary period would then not be shorter than 24\,yrs. However, other evidence for binarity is lacking in available data. The object \cygd, therefore, appears as a complex system, requiring additional monitoring, notably to search for binarity signatures.

\section{Conclusions}
\label{section_conclusions}
Using a dedicated X-ray monitoring, we have analyzed the behavior of three massive stars in the Cyg\,OB2 association: \cygc, \cygh, and \cygd.

First, we have enlarged the orbital coverage of \cygh, and confirmed the previous results reported for the star: a slight temperature increase at apastron, an additional absorption maximum at periastron and minimum at apastron, and a flux maximum at an intermediate phase ($\phi\sim0.8$). We have further analyzed the variability of the individual normalization factors, showing that they agree with expectations for WWCs. With data covering $\sim8.2$\,yrs (more than 100 cycles), the phase-locked variations of the system are now clearly ascertained, and the predicted hysteresis behavior of the fluxes is detected for the first time in a colliding wind binary. While observations are sufficient in number for covering the whole orbit, a dedicated full hydro model would be required to improve our understanding of the collision.

Second, we have reanalyzed the X-ray emission of \cygc\ using more data, showing that a modulation with the short period of the binary is unlikely. The most probable explanation is a modulation linked to the 6.7\,yr period of the tertiary, which explains the flux and absorption variations well. However, for in-depth testing, a precise ephemeris of the third star is needed, which is not (yet) available. Light travel time delaying the eclipses and radial velocity variations might help ascertain the orbit, and we have performed the first steps in this direction.

Third, our data shed a new light on the X-ray emission of \cygd, a peculiar B-hypergiant. The overall flux has been decreasing over the last decade, which could be associated with changes in a wind-wind collision occurring in a (very) long-period binary or with the aftermath of an eruption. Further monitoring is now required to pinpoint the nature of this peculiar object.

\begin{acknowledgements}
We thank the \sw\ PI, Dr Neil Gehrels, and the \xmm\ project scientist, Norbert Schartel, for having made the X-ray monitoring possible. We also thank the referee for his suggestions that helped improve the paper. YN acknowledges useful discussion with Kim Page on \sw\ data reduction and calibration. The Li\`ege team acknowledges support from the Fonds National de la Recherche Scientifique (Belgium), the Communaut\'e Fran\c caise de Belgique, the PRODEX XMM and Integral contracts, and the `Action de Recherche Concert\'ee' (CFWB-Acad\'emie Wallonie Europe). ADS and CDS were used for preparing this document. 
\end{acknowledgements}

%-------------------------------------------------------------------

\end{document}